# On Equivalence and Canonical Forms in the LF Type Theory

Robert Harper and Frank Pfenning
October 26, 2018
CMU-CS-00-148, Revised

School of Computer Science
Carnegie Mellon University
Pittsburgh, PA 15213

## Abstract

Decidability of definitional equality and conversion of terms into canonical form play a central role in the meta-theory of a type-theoretic logical framework. Most studies of definitional equality are based on a confluent, strongly-normalizing notion of reduction. Coquand has considered a different approach, directly proving the correctness of a practical equivalence algorithm based on the shape of terms. Neither approach appears to scale well to richer languages with unit types or subtyping, and neither directly addresses the problem of conversion to canonical form.

In this paper we present a new, type-directed equivalence algorithm for the LF type theory that overcomes the weaknesses of previous approaches. The algorithm is practical, scales to more expressive languages, and yields a new notion of canonical form sufficient for adequate encodings of logical systems. The algorithm is proved complete by a Kripke-style logical relations argument similar to that suggested by Coquand. Crucially, both the algorithm itself and the logical relations rely only on the shapes of types, ignoring dependencies on terms.

This work was sponsored NSF Grant CCR-9619584.

The views and conclusions contained in this document are those of the authors and should not be interpreted as representing the official policies, either expressed or implied, of NSF or the U.S. Government.

# Contents

# 1  Introduction

At present the mechanization of constructive reasoning relies almost entirely on type theories of various forms. The principal reason is that the computational meaning of constructive proofs is an integral part of the type theory itself. The main computational mechanism in such type theories is reduction, which has therefore been studied extensively.

For logical frameworks the case for type theoretic meta-languages is also compelling, since they allow us to internalize deductions as objects. The validity of a deduction is then verified by type-checking in the meta-language. To ensure that proof checking remains decidable under this representation, the type checking problem for the meta-language must also be decidable. To support deductive systems of practical interest, the type theory must support *dependent types*, that is, types that depend on objects.

The correctness of the representation of a logic in type theory is given by an *adequacy theorem* that correlates the syntax and deductions of the logic with *canonical forms* of suitable type. To establish a precise correspondence, canonical forms are taken to be $\beta$-normal, $\eta$-long forms. In particular, it is important that canonical forms enjoy the property that constants and variables of higher type are "fully applied" — that is, each occurrence is applied to enough arguments to reach a base type.

Thus we see that the methodology of logical frameworks relies on two fundamental meta-theoretic results: the decidability of type checking, and the existence of canonical forms. For many type theories the decidability of type checking is easily seen to reduce to the decidability of definitional equality of types and terms. Therefore we focus attention on the decision problem for definitional equality and on the conversion of terms to canonical form.

Traditionally, both problems have been treated by considering normal forms for $\beta$, and possibly $\eta$, reduction. If we take definitional equality to be conversion, then its decidability follows from confluence and strong normalization for the corresponding notion of reduction. In the case of $\beta$-reduction this approach to deciding definitional equality works well, but for $\beta\eta$-reduction the situation is much more complex. In particular, $\beta\eta$-reduction is confluent only for well-typed terms, and subject reduction depends on strengthening, which is difficult to prove directly.

These technical problems with $\beta\eta$-reduction have been addressed in work by Salvesen [Sal90], Geuvers [Geu92] and later with a different method by Goguen [Gog99], but nevertheless several problems remain. First, canonical forms are not $\beta\eta$-normal forms and so conversion to canonical form must be handled separately. The work by Dowek et al. [DHW93] shows how to do this for the Calculus of Constructions, but it is not clear that their approach would scale to richer theories such as those including linear types, unit types, or subtyping. Second, the algorithms implicit in the reduction-based accounts are not practical; if two terms are *not* definitionally equal, we can hope to discover this without reducing both to normal form.

These problems were side-stepped in the original paper on the LF logical framework [HHP93] by restricting attention to $\beta$-conversion for definitional equality. This is sufficient if we also restrict attention to $\eta$-long forms [FM90, Cer96]. This restriction is somewhat unsatisfactory, especially in linear variants of LF [CP98].

More recently, $\eta$-expansion has been studied in its own right, using modification of standard techniques from rewriting theory to overcome the lack of strong normalization when expansion is not restricted [JG95, Gha97]. In the dependently typed case, even the definition of long normal form is not obvious [DHW93] and the technical development is fraught with difficulties. We have not been able to reconstruct the proofs in [Gha97] and the development in [Vir99] relies on a complex intermediate system with annotated terms.

To address the problems of practicality, Coquand suggested abandoning reduction-based treatments of definitional equality in favor of a direct presentation of a practical equivalence algorithm [Coq91]. Coquand's approach is based on analyzing the "shapes" of terms, building in the principle of extensionality instead of relying on $\eta$-reduction or expansion. This algorithm improves on reduction-based approaches by avoiding explicit computation of normal forms, and allowing for early termination in the case that two terms are determined to be inequivalent. However, Coquand's approach can not be easily extended to richer type theories such as those with unit types. The problem can be traced to the reliance on the shape of terms, rather than on their classifying types, to guide the algorithm. For example, if $x$ and $y$ are two variables of unit type, they are definitionally equal, but structurally distinct. Moreover, their canonical forms would be the sole element of unit type. More recently, Compangnoni and Goguen [CG99] have developed an equality algorithm based on weak head-normal forms using typed operational semantics for a system with bounded operator abstraction. It is plausible that their method would also apply to LF, but, again, type theories with a less tractable notion of equality are likely to present problems.

In this paper we present a new type-directed algorithm for testing equality for a dependent type theory in the presence of $\beta$ and $\eta$-conversion, which generalizes the algorithm for the simply-typed case in [Pfe92]. We prove its correctness directly via logical relations. The essential idea is that we can erase dependencies when defining the logical relation, even though the domain of the relation contains dependently typed terms. This makes the definition obviously well-founded. Moreover, it means that the type-directed equality-testing algorithm on dependently typed terms requires only simple types. Consequently, transitivity of the algorithm is an easy property, which we were unable to obtain without this simplifying step. Soundness and completeness of the equality-testing algorithm yields the decidability of the type theory rather directly.

Another advantage of our approach is that it can be easily adapted to support adequacy proofs using a new notion of *quasi-canonical forms*, that is, canonical forms without type labels on $\lambda$-abstractions. We show that quasi-canonical forms of a given type are sufficient to determine the meaning of an object, since the type labels can be reconstructed (up to definitional equality) from the classifying type. Interestingly, recent research on dependently typed rewriting [Vir99] has also isolated equivalence classes of terms modulo conversion of the type labels as a critical concept. In some of the original work on Martin-Löf type theory [NPS90] and some subsequent studies [Str91], type theories without type labels have been studied, but to our knowledge they have not been considered with respect to bi-directional type-checking or adequacy proofs in logical framework representations.

There is now significant evidence that our construction is robust with respect to extension of the type theory with products, unit, linearity, subtyping and similar complicating factors. The reason is the flexibility of type-directed equality in the simply-typed case and the harmony between the definition of the logical relation and the algorithm, both of which are based on the erased types. The first author and Stone [SH00] have concurrently developed a variant of the technique presented here to handle a form of subtyping and singleton kinds. A number of papers subsequent to the original technical report describing our construction [HP99] have clearly demonstrated that the proposed technique is widely applicable. Vanderwaart and Crary [VC01] have adapted the ideas with minor modifications to give a proof of the decidability for linear LF that is stronger than the original one [CP98] since it does not require $\eta$-long forms from the start. The further adaptation to the case of an ordered linear type theory [Pol01] provides further evidence. Finally, the second author has adapted the technique to prove decidability and existence of canonical forms for a type theory with an internal notion of proof irrelevance and intensional types [Pfe01]. We conclude that our technique is directly applicable for a large class of dependent type theories where equality at the level of types is directly inherited from equality at the level of objects.

Despite this robustness for a whole class of extensions of the LF type theory, there are likely to be difficulties in applying our techniques in the impredicative setting, or even in the case of predicative universes. It is essential to our method that injectivity of products can be proved without first proving subject reduction and a Church-Rosser theorem; the reverse is the case for pure type systems [Bar92, Geu92].

More generally, it is not clear how to apply our ideas when faced with a complex notion of equality at the level of types unless it is directly inherited from the level of objects. Our formulation of LF omits type-level $\lambda$-abstraction precisely so we can prove injectivity of products at an early stage. Note that this is not a restriction from the point of view of our applications: Geuvers and Barendsen [GB99] have shown that LF without family level $\lambda$-abstraction is just as expressive as full LF. However, Vanderwaart and Crary [VC01] have shown that Coquand's technique for handling type-level $\lambda$-abstractions can be adapted to our proof by carrying out a separate, second logical relations argument. We suspect that this may be extended to the case of predicative universes, but the impredicative case is likely to require completely new ideas as discussed in the conclusion.

Our approach is similar to the technique of typed operational semantics of Goguen [Gog94, Gog99] in that both take advantage of types during reduction. However, as pointed out by Goguen [Gog99], the development of the complete meta-theory of the LF requires the use of an untyped reduction relation. Our techniques avoid this entirely, fulfilling Goguen's conjecture that a complete development should be possible without resorting to untyped methods.

The remainder of the paper is organized as follows. In Section 2 we present a variant of the LF type theory and investigate its elementary syntactic properties. It can be seen to be equivalent to the original LF proposal with $\beta\eta$-conversion at the end of our development. In Section 3 we present an algorithm for testing equality that uses an approximate typing relation and exploits extensionality. In Section 4 we show that the algorithm is complete via a Kripke-logical relation argument using approximate types. This is complemented by a corresponding soundness proof for the algorithm on well-typed terms in Section 5. In Section 6 we exploit the soundness and completeness of the algorithm to obtain decidability for all judgments of the LF type theory with an extensional equality. In Section 7 we show how to extract quasi-canonical forms from our conversion algorithm. They differ from long $\beta\eta$-normal forms in that object carry not type labels. We show that this is sufficient for adequacy theorems in the logical framework since such type labels are determined uniquely modulo definitional equality. In the conclusion in Section 8 we discuss some possible limitations of our technique and mention some further work.

## 2  A Variant of the LF Type Theory

Syntactically, our formulation of the LF type theory follows the original proposal by Harper, Honsell and Plotkin [HHP93], except that we omit type-level $\lambda$-abstraction. This simplifies the proof of the soundness theorem considerably, since we can prove the injectivity of products (Lemma 12) at an early stage. In practice, this restriction has no impact since types in normal form never contain type-level $\lambda$-abstractions. This observation has been formalized by Geuvers and Barendsen [GB99].

## 2.1 Syntax

$$
\begin{aligned}
\text{Kinds} \quad K &::= \text{type} \mid \Pi x{:}A.\ K \\
\text{Families} \quad A &::= a \mid A\ M \mid \Pi x{:}A_1.\ A_2 \\
\text{Objects} \quad M &::= c \mid x \mid \lambda x{:}A.\ M \mid M_1\ M_2 \\
\text{Signatures} \quad \Sigma &::= \cdot \mid \Sigma, a{:}K \mid \Sigma, c{:}A \\
\text{Contexts} \quad \Gamma &::= \cdot \mid \Gamma, x{:}A
\end{aligned}
$$

We use $K$ for kinds, $A, B, C$ for type families, $M, N, O$ for objects, $\Gamma, \Psi$ for contexts and $\Sigma$ for signatures. We also use the symbol "kind" to classify the valid kinds. We consider terms that differ only in the names of their bound variables as identical. We write $[N/x]M$, $[N/x]A$ and $[N/x]K$ for capture-avoiding substitution. Signatures and contexts may declare each constant and variable at most once. For example, when we write $\Gamma, x{:}A$ we assume that $x$ is not already declared in $\Gamma$. If necessary, we tacitly rename $x$ before adding it to the context $\Gamma$.

## 2.2 Substitutions

In the logical relations argument, we require a notion of simultaneous substitution.

$$
\text{Substitutions} \quad \sigma ::= \cdot \mid \sigma, M/x
$$

We assume that no variable is defined more than once in any substitution which can be achieved by appropriate renaming where necessary. We do not develop a notion and theory of well-typed substitutions, since it is unnecessary for our purposes. However, when applying a substitution $\sigma$ to a term $M$ we maintain the invariant that all free variables in $M$ occur in the domain of $\sigma$, and similarly for families and kinds.

We write $\text{id}_\Gamma$ for the identity substitution on the context $\Gamma$. We use the notation $M[\sigma]$, $A[\sigma]$ and $K[\sigma]$ for the *simultaneous* substitution by $\sigma$ into an object, family, or kind. It is defined by simultaneous induction on the structure of objects, families, and kinds.

$$
\begin{aligned}
x[\sigma] &= M \quad \text{where } M/x \text{ in } \sigma \\
c[\sigma] &= c \\
(\lambda x{:}A.\ M)[\sigma] &= \lambda x{:}A[\sigma].\ M[\sigma, x/x] \\
(M\ N)[\sigma] &= M[\sigma]\ N[\sigma] \\
\\
a[\sigma] &= a \\
(A\ M)[\sigma] &= A[\sigma]\ M[\sigma] \\
(\Pi x{:}A.\ B)[\sigma] &= \Pi x{:}A[\sigma].\ B[\sigma, x/x] \\
\\
\text{type}[\sigma] &= \text{type} \\
(\Pi x{:}A.\ K)[\sigma] &= \Pi x{:}A[\sigma].\ K[\sigma, x/x]
\end{aligned}
$$

Extending the substitution $\sigma$ to $(\sigma, x/x)$ may require some prior renaming of the variable $x$ in order to satisfy our assumption on substitutions.

## 2.3 Judgments

The LF type theory is defined by the following judgments.

$$
\begin{array}{ll}
\vdash \Sigma \text{ sig} & \Sigma \text{ is a valid signature} \\
\vdash_{\Sigma} \Gamma \text{ ctx} & \Gamma \text{ is a valid context} \\
\Gamma \vdash_{\Sigma} M : A & M \text{ has type } A \\
\Gamma \vdash_{\Sigma} A : K & A \text{ has type } K \\
\Gamma \vdash_{\Sigma} K : \text{kind} & K \text{ is a valid kind} \\
\Gamma \vdash_{\Sigma} M = N : A & M \text{ equals } N \text{ at type } A \\
\Gamma \vdash_{\Sigma} A = B : K & A \text{ equals } B \text{ at kind } K \\
\Gamma \vdash_{\Sigma} K = L : \text{kind} & K \text{ equals } L
\end{array}
$$

For the judgment $\vdash_{\Sigma} \Gamma$ ctx we presuppose that $\Sigma$ is a valid signature. For the remaining judgments of the form $\Gamma \vdash_{\Sigma} J$ we presuppose that $\Sigma$ is a valid signature and that $\Gamma$ is valid in $\Sigma$. For the sake of brevity we omit the signature $\Sigma$ from all judgments but the first, since it does not change throughout a derivation.

If $J$ is a typing or equality judgment, then we write $J[\sigma]$ for the obvious substitution of $J$ by $\sigma$. For example, if $J$ is $M : A$, then $J[\sigma]$ stands for the judgment $M[\sigma] : A[\sigma]$.

## 2.4 Typing Rules

Our formulation of the typing rules is similar to the second version given in [HHP93]. In preparation for the various algorithms we presuppose and inductively preserve the validity of contexts involved in the judgments, instead of checking these properties at the leaves. This is a matter of expediency rather than necessity.

**Signatures**

$$
\dfrac{}{\vdash \cdot \text{ sig}} \qquad \dfrac{\vdash \Sigma \text{ sig} \quad \cdot \vdash_{\Sigma} K : \text{kind}}{\vdash \Sigma, a{:}K \text{ sig}} \qquad \dfrac{\vdash \Sigma \text{ sig} \quad \cdot \vdash_{\Sigma} A : \text{type}}{\vdash \Sigma, c{:}A \text{ sig}}
$$

From now on we fix a valid signature $\Sigma$ and omit it from the judgments.

**Contexts**

$$
\dfrac{}{\vdash \Gamma \text{ ctx}} \qquad \dfrac{\vdash \Gamma \text{ ctx} \quad \Gamma \vdash A : \text{type}}{\vdash \Gamma, x{:}A \text{ ctx}}
$$

From now on we presuppose that all contexts in judgments are valid, instead of checking it explicitly. This means, for example, that we have to verify the validity of the type labels in $\lambda$-abstractions before adding them to the context.

**Objects**

$$\frac{x{:}A \text{ in } \Gamma}{\Gamma \vdash x : A} \qquad \frac{c{:}A \text{ in } \Sigma}{\Gamma \vdash c : A}$$

$$\frac{\Gamma \vdash M_1 : \Pi x{:}A_2.\ A_1 \qquad \Gamma \vdash M_2 : A_2}{\Gamma \vdash M_1\ M_2 : [M_2/x]A_1}$$

$$\frac{\Gamma \vdash A_1 : \text{type} \qquad \Gamma, x{:}A_1 \vdash M_2 : A_2}{\Gamma \vdash \lambda x{:}A_1.\ M_2 : \Pi x{:}A_1.\ A_2}$$

$$\frac{\Gamma \vdash M : A \qquad \Gamma \vdash A = B : \text{type}}{\Gamma \vdash M : B}$$

**Families**

$$\frac{a{:}K \text{ in } \Sigma}{\Gamma \vdash a : K} \qquad \frac{\Gamma \vdash A : \Pi x{:}B.\ K \qquad \Gamma \vdash M : B}{\Gamma \vdash A\ M : [M/x]K}$$

$$\frac{\Gamma \vdash A_1 : \text{type} \qquad \Gamma, x{:}A_1 \vdash A_2 : \text{type}}{\Gamma \vdash \Pi x{:}A_1.\ A_2 : \text{type}}$$

$$\frac{\Gamma \vdash A : K \qquad \Gamma \vdash K = L : \text{kind}}{\Gamma \vdash A : L}$$

**Kinds**

$$\frac{}{\Gamma \vdash \text{type} : \text{kind}} \qquad \frac{\Gamma \vdash A : \text{type} \qquad \Gamma, x{:}A \vdash K : \text{kind}}{\Gamma \vdash \Pi x{:}A.\ K : \text{kind}}$$

## 2.5 Definitional Equality

The rules for definitional equality are written with the presupposition that a valid signature $\Sigma$ is fixed and that all contexts $\Gamma$ are valid. The intent is that equality implies validity of the objects, families, or kinds involved (see Lemma 7). In contrast to the original formulation in [HHP93], equality is based on a notion of parallel conversion plus extensionality, rather then $\beta\eta$-conversion. We believe this is a robust foundation, easily transferred to richer and more complicated type theories. Parallel conversion allows the equality judgment to be relatively independent from the typing judgment, thereby simplifying the completeness proof of our algorithm. It does not otherwise appear to be essential. The use of extensionality on the other hand is central.

Characteristically for parallel conversion, reflexivity is admissible (Lemma 2) which significantly simplifies the completeness proof for the algorithm to check equality. We enclose the admissible rules are in [brackets]. Some of the typing premises in the rules are redundant, but for technical reasons we cannot prove this until validity has been established. Such premises are enclosed in {braces}. Alternatively, it may be sufficient to check validity of the contexts at the leaves of the derivations (the cases for variables and constants), a technique used both in the original presentation of LF [HHP93] and Pure Type Systems [Bar92].

**Simultaneous Congruence**

$$\frac{x{:}A \text{ in } \Gamma}{\Gamma \vdash x = x : A} \qquad \frac{c{:}A \text{ in } \Sigma}{\Gamma \vdash c = c : A}$$

$$\frac{\Gamma \vdash M_1 = N_1 : \Pi x{:}A_2.\ A_1 \qquad \Gamma \vdash M_2 = N_2 : A_2}{\Gamma \vdash M_1\ M_2 = N_1\ N_2 : [M_2/x]A_1}$$

$$\frac{\Gamma \vdash A_1' = A_1 : \text{type} \qquad \Gamma \vdash A_1'' = A_1 : \text{type} \qquad \Gamma, x{:}A_1 \vdash M_2 = N_2 : A_2}{\Gamma \vdash \lambda x{:}A_1'.\ M_2 = \lambda x{:}A_1''.\ N_2 : \Pi x{:}A_1.\ A_2}$$

**Extensionality**

$$\frac{\Gamma \vdash A_1 : \text{type} \quad \{\Gamma \vdash M : \Pi x{:}A_1.\ A_2\} \quad \{\Gamma \vdash N : \Pi x{:}A_1.\ A_2\} \quad \Gamma, x{:}A_1 \vdash M\ x = N\ x : A_2}{\Gamma \vdash M = N : \Pi x{:}A_1.\ A_2}$$

**Parallel Conversion**

$$\frac{\{\Gamma \vdash A_1 : \text{type}\} \qquad \Gamma, x{:}A_1 \vdash M_2 = N_2 : A_2 \qquad \Gamma \vdash M_1 = N_1 : A_1}{\Gamma \vdash (\lambda x{:}A_1.\ M_2)\ M_1 = [N_1/x]N_2 : [M_1/x]A_2}$$

**Equivalence**

$$\frac{\Gamma \vdash M = N : A}{\Gamma \vdash N = M : A} \qquad \frac{\Gamma \vdash M = N : A \qquad \Gamma \vdash N = O : A}{\Gamma \vdash M = O : A}$$

$$\left[\ \frac{\Gamma \vdash M : A}{\Gamma \vdash M = M : A}\ \right]$$

**Type Conversion**

$$\frac{\Gamma \vdash M = N : A \qquad \Gamma \vdash A = B : \text{type}}{\Gamma \vdash M = N : B}$$

**Family Congruence**

$$\frac{a{:}K \text{ in } \Sigma}{\Gamma \vdash a = a : K}$$

$$\frac{\Gamma \vdash A = B : \Pi x{:}C.\ K \qquad \Gamma \vdash M = N : C}{\Gamma \vdash A\ M = B\ N : [M/x]K}$$

$$\frac{\Gamma \vdash A_1 = B_1 : \text{type} \qquad \{\Gamma \vdash A_1 : \text{type}\} \qquad \Gamma, x{:}A_1 \vdash A_2 = B_2 : \text{type}}{\Gamma \vdash \Pi x{:}A_1.\ A_2 = \Pi x{:}B_1.\ B_2 : \text{type}}$$

**Family Equivalence**

$$\frac{\Gamma \vdash A = B : K}{\Gamma \vdash B = A : K} \qquad \frac{\Gamma \vdash A = B : K \quad \Gamma \vdash B = C : K}{\Gamma \vdash A = C : K}$$

$$\left[ \frac{\Gamma \vdash A : K}{\Gamma \vdash A = A : K} \right]$$

**Kind Conversion**

$$\frac{\Gamma \vdash A = B : K \quad \Gamma \vdash K = L : \text{kind}}{\Gamma \vdash A = B : L}$$

**Kind Congruence**

$$\overline{\Gamma \vdash \text{type} = \text{type} : \text{kind}}$$

$$\frac{\Gamma \vdash A = B : \text{type} \quad \{\Gamma \vdash A : \text{type}\} \quad \Gamma, x{:}A \vdash K = L : \text{kind}}{\Gamma \vdash \Pi x{:}A.\ K = \Pi x{:}B.\ L : \text{kind}}$$

**Kind Equivalence**

$$\frac{\Gamma \vdash K = L : \text{kind}}{\Gamma \vdash L = K : \text{kind}} \qquad \frac{\Gamma \vdash K = L : \text{kind} \quad \Gamma \vdash L = L' : \text{kind}}{\Gamma \vdash K = L' : \text{kind}}$$

$$\left[ \frac{\Gamma \vdash K : \text{kind}}{\Gamma \vdash K = K : \text{kind}} \right]$$

## 2.6 Elementary Properties of Typing and Definitional Equality

We establish some elementary properties of the judgments pertaining to the interpretation of contexts. There is an alternative route to these properties by first introducing a notion of substitution and well-typed substitution.

First we establish weakening for all judgments of the type theory. We use $J$ to stand for any of the relevant judgments of the type theory in order to avoid repetitive statements. We extend the notation of substitution to all judgments of the type theory in the obvious way. For example, if $J$ is $N : B$ then $[M/x]J$ is $[M/x]N : [M/x]B$.

**Lemma 1 (Weakening)** *If $\Gamma, \Gamma' \vdash J$ then $\Gamma, x{:}A, \Gamma' \vdash J$.*

**Proof:** By straightforward induction over the structure of the given derivation. □

Note that exchange for independent hypotheses and contraction are also admissible, but we elide the statement of these properties here since they are not needed for the results in this paper. Next we show that reflexivity is admissible.

**Lemma 2 (Reflexivity)**

1. *If $\Gamma \vdash M : A$ then $\Gamma \vdash M = M : A$.*

2. *If $\Gamma \vdash A : K$ then $\Gamma \vdash A = A : K$.*

3. *If $\Gamma \vdash K :$ kind *then* $\Gamma \vdash K = K :$ kind.*

**Proof:** By induction over the structure of the given derivations. In each case the result follows immediately from the available induction hypotheses.

□

Next we prove the central substitution property.

**Lemma 3 (Substitution Property for Typing and Definitional Equality)**
*Assume $\Gamma, x{:}A, \Gamma'$ is a valid context. If $\Gamma \vdash M : A$ and $\Gamma, x{:}A, \Gamma' \vdash J$ then $\Gamma, [M/x]\Gamma' \vdash [M/x]J$.*

**Proof:** By straightforward inductions over the structure of the given derivations. □

The next lemma applies in a number of the proofs in the remainder of this section.

**Lemma 4 (Context Conversion)** *Assume $\Gamma, x{:}A$ is a valid context and $\Gamma \vdash B :$ type. If $\Gamma, x{:}A \vdash J$ and $\Gamma \vdash A = B :$ type *then* $\Gamma, x{:}B \vdash J$.*

**Proof:** Direct, taking advantage of weakening and substitution.

| | |
|---|---|
| $\Gamma, x{:}B \vdash x : B$ | By rule (variable) |
| $\Gamma \vdash B = A :$ type | By symmetry from assumption |
| $\Gamma, x{:}B \vdash x : A$ | By rule (type conversion) |
| $\Gamma, x'{:}A \vdash [x'/x]J$ | By renaming from assumption |
| $\Gamma, x{:}B, x'{:}A \vdash [x'/x]J$ | By weakening |
| $\Gamma, x{:}B \vdash [x/x'][x'/x]J$ | By substitution property |
| $\Gamma, x{:}B \vdash J$ | By definition of substitution |

□

Besides substitution, we require functionality for the typing judgments. Note that a stronger version of functionality for equality judgments must be postponed until validity (Lemma 7) has been proven. We state this in a slightly more general form than required below in order to prove it inductively.

**Lemma 5 (Functionality for Typing)** *Assume $\Gamma, x{:}A, \Gamma'$ is a valid context, $\Gamma \vdash M = N : A$, $\Gamma \vdash M : A$, and $\Gamma \vdash N : A$.*

1. *If $\Gamma, x{:}A, \Gamma' \vdash P : B$ then $\Gamma, [M/x]\Gamma' \vdash [M/x]P = [N/x]P : [M/x]B$.*

2. *If $\Gamma, x{:}A, \Gamma' \vdash B : K$ then $\Gamma, [M/x]\Gamma' \vdash [M/x]B = [N/x]B : [M/x]K$.*

3. *If $\Gamma, x{:}A, \Gamma' \vdash K :$ kind *then* $\Gamma, [M/x]\Gamma' \vdash [M/x]K = [N/x]K :$ kind.*

**Proof:** By a straightforward induction on the given derivation $\mathcal{D}$ in each case. We show some representative cases.

**Case:**

$$\mathcal{D} = \overline{\Gamma, x{:}A, \Gamma' \vdash x : A}$$

$$\Gamma \vdash M = N : A \qquad \text{Assumption}$$
$$\Gamma, [M/x]\Gamma' \vdash M = N : A \qquad \text{By weakening}$$

**Case:**

$$\mathcal{D} = \frac{y{:}B \text{ in } \Gamma \text{ or } \Gamma'}{\Gamma, x{:}A, \Gamma' \vdash y : B}$$

$y{:}B$ in $\Gamma$ or $y{:}[M/x]B$ in $[M/x]\Gamma'$      By definition of substitution
$\Gamma, [M/x]\Gamma' \vdash y = y : [M/x]B$     By rule

**Case:**

$$\mathcal{D} = \frac{\begin{array}{cc}\mathcal{D}_1 & \mathcal{D}_2\\ \Gamma, x{:}A, \Gamma' \vdash P_1 : \Pi y{:}B_2.\ B_1 & \Gamma, x{:}A, \Gamma' \vdash P_2 : B_2\end{array}}{\Gamma, x{:}A, \Gamma' \vdash P_1\ P_2 : [P_2/y]B_1}$$

$\Gamma, [M/x]\Gamma' \vdash [M/x]P_1 = [N/x]P_1 : \Pi y{:}[M/x]B_2.\ [M/x]B_1$     By i.h. on $\mathcal{D}_1$
$\Gamma, [M/x]\Gamma' \vdash [M/x]P_2 = [N/x]P_2 : [M/x]B_2$     By i.h. on $\mathcal{D}_2$
$\Gamma, [M/x]\Gamma' \vdash ([M/x]P_1)\,([M/x]P_2) = ([N/x]P_1)\,([N/x]P_2) : [([M/x]P_2)/y]([M/x]B_1)$     By rule
$\Gamma, [M/x]\Gamma' \vdash [M/x](P_1\ P_2) = [N/x](P_1\ P_2) : [M/x]([P_2/y]B_1)$     By properties of substitution

**Case:**

$$\mathcal{D} = \frac{\begin{array}{cc}\mathcal{D}_1 & \mathcal{D}_2\\ \Gamma, x{:}A, \Gamma' \vdash B_1 : \text{type} & \Gamma, x{:}A, \Gamma', y{:}B_1 \vdash P_2 : B_2\end{array}}{\Gamma, x{:}A, \Gamma' \vdash \lambda y{:}B_1.\ P_2 : \Pi y{:}B_1.\ B_2}$$

$\Gamma, [M/x]\Gamma' \vdash [M/x]B_1 = [N/x]B_1 : \text{type}$     By i.h. on $\mathcal{D}_1$
$\Gamma, [M/x]\Gamma, y{:}[M/x]B_1 \vdash [M/x]P_2 = [N/x]P_2 : [M/x]B_2$     By i.h. on $\mathcal{D}_2$
$\Gamma, [M/x]\Gamma \vdash [M/x]B_1 : \text{type}$     By substitution property
$\Gamma, [M/x]\Gamma \vdash [M/x]B_1 = [M/x]B_1 : \text{type}$     By reflexivity
$\Gamma, [M/x]\Gamma \vdash [N/x]B_1 = [M/x]B_1 : \text{type}$     By symmetry
$\Gamma, [M/x]\Gamma \vdash \lambda y{:}[M/x]B_1.\ P_2 = \lambda y{:}[N/x]B_1.\ [N/x]P_2 : \Pi y{:}[M/x]B_1.\ [M/x]B_2$     By rule

**Case:**

$$\mathcal{D} = \frac{\begin{array}{cc}\mathcal{D}_1 & \mathcal{D}_2\\ \Gamma, x{:}A, \Gamma' \vdash P : C & \Gamma, x{:}A, \Gamma' \vdash C = B : \text{type}\end{array}}{\Gamma, x{:}A, \Gamma' \vdash P : B}$$

$\Gamma, [M/x]\Gamma' \vdash [M/x]P = [N/x]P : [M/x]C$     By i.h. on $\mathcal{D}_1$
$\Gamma, [M/x]\Gamma' \vdash [M/x]C = [M/x]B : \text{type}$     By substitution property
$\Gamma, [M/x]\Gamma' \vdash [M/x]P = [N/x]P : [M/x]B$     By rule (type conversion)

$\square$

We have to postpone the general inversion properties until validity (Lemma 7) has been proven. However, we need the simpler property of inversion on products in order to prove validity.

**Lemma 6 (Inversion on Products)**

1. *If $\Gamma \vdash \Pi x{:}A_1.\ A_2 : K$ then $\Gamma \vdash A_1 : $ type, and $\Gamma, x{:}A_1 \vdash A_2 : $ type.*

2. *If $\Gamma \vdash \Pi x{:}A.\ K : $ kind then $\Gamma \vdash A : $ type and $\Gamma, x{:}A \vdash K : $ kind.*

**Proof:** Part (1) follows by induction on the given derivation since it is stated for general kinds $K$. Part (2) is immediate by inversion. □

Now we have the necessary properties to prove the critical validity property. Recall our general assumption that all signatures are valid.

**Lemma 7 (Validity)** *Assume $\Gamma$ is a valid context.*

1. *If $\Gamma \vdash M : A$ then $\Gamma \vdash A : $ type.*

2. *If $\Gamma \vdash M = N : A$, then $\Gamma \vdash M : A$, $\Gamma \vdash N : A$, and $\Gamma \vdash A : $ type.*

3. *If $\Gamma \vdash A : K$, then $\Gamma \vdash K : $ kind.*

4. *If $\Gamma \vdash A = B : K$, then $\Gamma \vdash A : K$, $\Gamma \vdash B : K$, and $\Gamma \vdash K : $ kind.*

5. *If $\Gamma \vdash K = L : $ kind, then $\Gamma \vdash K : $ kind and $\Gamma \vdash L : $ kind.*

**Proof:** By a straightforward simultaneous induction on derivations. Functionality for typing (Lemma 5) is required to handle the case of applications. The typing premises on the rule of extensionality ensure that strengthening is not required.

**Case:**

$$\mathcal{E} = \frac{\begin{array}{c}\mathcal{E}_1\\ \Gamma \vdash M_1 = N_1 : \Pi x{:}A_2.\ A_1\end{array} \qquad \begin{array}{c}\mathcal{E}_2\\ \Gamma \vdash M_2 = N_2 : A_2\end{array}}{\Gamma \vdash M_1\ M_2 = N_1\ N_2 : [M_2/x]A_1}$$

| | |
|---|---:|
| $\Gamma \vdash M_1 : \Pi x{:}A_2.\ A_1$ | |
| $\Gamma \vdash N_1 : \Pi x{:}A_2.\ A_1$ | |
| $\Gamma \vdash \Pi x{:}A_2.\ A_1 : $ type | By i.h. on $\mathcal{E}_1$ |
| $\Gamma \vdash M_2 : A_2$ | |
| $\Gamma \vdash N_2 : A_2$ | |
| $\Gamma \vdash A_2 : $ type | By i.h. on $\mathcal{E}_2$ |
| $\Gamma, x{:}A_2 \vdash A_1 : $ type | By inversion on products (Lemma 6) |
| $\Gamma \vdash [M_2/x]A_1 : $ type | By substitution property |
| $\Gamma \vdash M_1\ M_2 : [M_2/x]A_1$ | By rule |
| $\Gamma \vdash N_1\ N_2 : [N_2/x]A_1$ | By rule |
| $\Gamma \vdash [M_2/x]A_1 = [N_2/x]A_1 : $ type | By functionality (Lemma 5) |
| $\Gamma \vdash N_1\ N_2 : [M_2/x]A_1$ | By symmetry and type conversion |

□

With the central validity property, we can show a few other syntactic results. The first of these is that functionality holds even for the equality judgments. Since this can be proven directly, we state it in the more restricted form in which it is needed subsequently.

**Lemma 8 (Functionality for Equality)** *Assume $\Gamma, x{:}A$ is a valid context and $\Gamma \vdash M = N : A$.*

1. *If $\Gamma, x{:}A \vdash O = P : B$ then $\Gamma \vdash [M/x]O = [N/x]P : [M/x]B$.*

2. *If $\Gamma, x{:}A \vdash B = C : K$ then $\Gamma \vdash [M/x]B = [N/x]C : [M/x]K$.*

3. *If $\Gamma, x{:}A \vdash K = L : \text{kind}$ then $\Gamma \vdash [M/x]K = [N/x]L : \text{kind}$.*

**Proof:** Direct, using validity, substitution, and functionality for typing. We show only the proof of part (1).

| | |
|---|---:|
| $\Gamma, x{:}A \vdash O = P : B$ | Assumption |
| $\Gamma \vdash M = N : A$ | Assumption |
| $\Gamma \vdash M : A$ | By validity |
| $\Gamma \vdash N : A$ | By validity |
| $\Gamma \vdash [M/x]O = [M/x]P : [M/x]B$ | By substitution |
| $\Gamma, x{:}A \vdash P : B$ | By validity |
| $\Gamma \vdash [M/x]P = [N/x]P : [M/x]B$ | By functionality for typing (Lemma 5) |
| $\Gamma \vdash [M/x]O = [N/x]P : [M/x]B$ | By rule (transitivity) |

At the level of objects it is also possible to derive functionality by introducing $\lambda$-abstractions, applications, and parallel conversion. However, this is not possible at the level of families, since there is no corresponding $\lambda$-abstraction. □

The second consequence of validity is a collection of inversion properties which generalize inversion of products (Lemma 6).

**Lemma 9 (Typing Inversion)** *Assume $\Gamma$ is a valid context.*

1. *If $\Gamma \vdash x : A$ then $x{:}B$ in $\Gamma$ and $\Gamma \vdash A = B : \text{type}$ for some $B$.*

2. *If $\Gamma \vdash c : A$ then $c{:}B$ in $\Gamma$ and $\Gamma \vdash A = B : \text{type}$ for some $B$.*

3. *If $\Gamma \vdash M_1 \, M_2 : A$ then $\Gamma \vdash M_1 : \Pi x{:}A_2.\, A_1$, $\Gamma \vdash M_2 : A_2$ and $\Gamma \vdash [M_2/x]A_1 = A : \text{type}$ for some $A_1$ and $A_2$.*

4. *If $\Gamma \vdash \lambda x{:}A.\, M : B$, then $\Gamma \vdash B = \Pi x{:}A.\, A' : \text{type}$, $\Gamma \vdash A : \text{type}$, and $\Gamma, x{:}A \vdash M : A'$.*

5. *If $\Gamma \vdash \Pi x{:}A_1.\, A_2 : K$ then $\Gamma \vdash K = \text{type} : \text{kind}$, $\Gamma \vdash A_1 : \text{type}$ and $\Gamma, x{:}A_1 \vdash A_2 : \text{type}$.*

6. *If $\Gamma \vdash a : K$, then $a{:}L$ in $\Sigma$ and $\Gamma \vdash K = L : \text{kind}$ for some $L$.*

7. *If $\Gamma \vdash A\, M : K$, then $\Gamma \vdash A : \Pi x{:}A_1.\, K_2$, $\Gamma \vdash M : A_1$, and $\Gamma \vdash K = [M/x]K_2 : \text{kind}$.*

8. *If $\Gamma \vdash \Pi x{:}A_1.\, K_2 : \text{kind}$, then $\Gamma \vdash A_1 : \text{type}$ and $\Gamma, x{:}A_1 \vdash K_2 : \text{kind}$.*

**Proof:** By a straightforward induction on typing derivations. Validity is needed in most cases in order to apply reflexivity. □

We can now show that some of the typing premises in the inference rules are redundant.

**Lemma 10 (Redundancy of Typing Premises)** *The indicated typing premises in the rules of parallel conversion, family congruence, and type congruence are redundant.*

**Proof:** Straightforward from validity. □

**Lemma 11 (Equality Inversion)** *Assume $\Gamma$ is a valid context.*

1. *If $\Gamma \vdash A = \Pi x{:}B_1.\, B_2 : \text{type}$ or $\Gamma \vdash \Pi x{:}B_1.\, B_2 = A : \text{type}$ then $A = \Pi x{:}A_1.\, A_2$ for some $A_1$ and $A_2$ such that $\Gamma \vdash A_1 = B_1 : \text{type}$ and $\Gamma, x{:}A_1 \vdash A_2 = B_2 : \text{type}$.*

2. *If $\Gamma \vdash K = \text{type} : \text{kind}$ or $\Gamma \vdash \text{type} = K : \text{kind}$ then $K = \text{type}$.*

3. *If $\Gamma \vdash K = \Pi x{:}B_1.\, L_2 : \text{kind}$ or $\Gamma \vdash \Pi x{:}B_1.\, L_2 = K : \text{kind}$ then $K = \Pi x{:}A_1.\, K_2$ such that $\Gamma \vdash A_1 = B_1 : \text{type}$ and $\Gamma, x{:}A_1 \vdash K_2 = L_2 : \text{kind}$.*

**Proof:** By induction on the given equality derivations. There are some subtle points in the proof of part 1, so we show two cases. Note that adding a family-level $\lambda$ would prevent proving this result at such an early stage.

**Case:**

$$\mathcal{E} = \frac{\overset{\mathcal{E}_1}{\Gamma \vdash A = C : \text{type}} \quad \overset{\mathcal{E}_2}{\Gamma \vdash C = \Pi x{:}B_1.\, B_2 : \text{type}}}{\Gamma \vdash A = \Pi x{:}B_1.\, B_2 : \text{type}}$$

$C = \Pi x{:}C_1.\, C_2$ for some $C_1$ and $C_2$ such that
$\Gamma \vdash C_1 = B_1 : \text{type}$ and
$\Gamma, x{:}C_1 \vdash C_2 = B_2 : \text{type}$                     By i.h. (1) on $\mathcal{E}_2$
$A = \Pi x{:}A_1.\, A_2$ for some $A_1$ and $A_2$ such that
$\Gamma \vdash A_1 = C_1 : \text{type}$ and
$\Gamma, x{:}A_1 \vdash A_2 = C_2 : \text{type}$                     By i.h. (1) on $\mathcal{E}_1$
$\Gamma \vdash A_1 = B_1 : \text{type}$                                       By rule (transitivity)
$\Gamma, x{:}A_1 \vdash C_2 = B_2 : \text{type}$                      By context conversion (Lemma 4)
$\Gamma, x{:}A_1 \vdash A_2 = B_2 : \text{type}$                          By rule (transitivity)

**Case:**

$$\mathcal{E} = \frac{\overset{\mathcal{E}_1}{\Gamma \vdash A = \Pi x{:}B_1.\, B_2 : K} \quad \overset{\mathcal{E}_2}{\Gamma \vdash K = \text{type} : \text{kind}}}{\Gamma \vdash A = \Pi x{:}B_1.\, B_2 : \text{type}}$$

$K = \text{type}$                                                   By i.h. (2) on $\mathcal{E}_2$
$A = \Pi x{:}A_1.\, A_2$ for some $A_1$ and $A_2$ such that
$\Gamma \vdash A_1 = B_1 : \text{type}$ and
$\Gamma, x{:}A_1 \vdash A_2 = B_2 : \text{type}$                     By i.h. (1) on $\mathcal{E}_1$

□

**Lemma 12 (Injectivity of Products)**

1. *If $\Gamma \vdash \Pi x{:}A_1.\, A_2 = \Pi x{:}B_1.\, B_2 : \text{type}$ then $\Gamma \vdash A_1 = B_1 : \text{type}$ and $\Gamma, x{:}A_1 \vdash A_2 = B_2 : \text{type}$.*

2. *If $\Gamma \vdash \Pi x{:}A_1.\, K_2 = \Pi x{:}B_1.\, L_2 : \text{kind}$ then $\Gamma \vdash A_1 = B_1 : \text{type}$ and $\Gamma, x{:}A_1 \vdash K_2 = L_2 : \text{kind}$.*

**Proof:** Immediate by equality inversion (Lemma 11). □

# 3  Algorithmic Equality

The algorithm for deciding equality can be summarized as follows:

1. When comparing objects at function type, apply extensionality.

2. When comparing objects at base type, reduce both sides to weak head-normal form and then compare heads directly and, if they are equal, each corresponding pair of arguments according to their type.

Since this algorithm is type-directed in case (1) we need to carry types. Unfortunately, this makes it difficult to prove correctness of the algorithm in the presence of dependent types, because transitivity is not an obvious property. The informal description above already contains a clue to the solution: we do not need to know the precise type of the objects we are comparing, as long as we know that they are functions.

We therefore define a calculus of simple types and an erasure function $()^-$ that eliminates dependencies for the purpose of this algorithm. The same idea is used later in the definition of the Kripke logical relation to prove completeness of the algorithm.

We write $\alpha$ to stands for simple base types and we have two special type constants, $\text{type}^-$ and $\text{kind}^-$, for the equality judgments at the level of types and kinds.

$$
\begin{array}{rcl}
\text{Simple Kinds} \quad \kappa & ::= & \text{type}^- \mid \tau \to \kappa \\
\text{Simple Types} \quad \tau & ::= & \alpha \mid \tau_1 \to \tau_2 \\
\text{Simple Contexts} \quad \Delta & ::= & \cdot \mid \Delta, x{:}\tau
\end{array}
$$

We use $\tau, \theta, \delta$ for simple types and $\Delta, \Theta$ for contexts declaring simple types for variables. We also use $\text{kind}^-$ in a similar role to kind in the LF type theory.

We write $A^-$ for the simple type that results from erasing dependencies in $A$, and similarly $K^-$. We translate each constant type family $a$ to a base type $a^-$ and extend this to all type families. We extend it further to contexts by applying it to each declaration.

$$
\begin{array}{rcl}
(a)^- & = & a^- \\
(A\,M)^- & = & A^- \\
(\Pi x{:}A_1.\ A_2)^- & = & A_1^- \to A_2^- \\
(\text{type})^- & = & \text{type}^- \\
(\Pi x{:}A.\ K)^- & = & A^- \to K^- \\
(\text{kind})^- & = & \text{kind}^- \\
(\cdot)^- & = & \cdot \\
(\Gamma, x{:}A)^- & = & \Gamma^-, x{:}A^-
\end{array}
$$

We need the property that the erasure of a type or kind remains invariant under equality and substitution.

**Lemma 13 (Erasure Preservation)**

1. *If* $\Gamma \vdash A = B : K$ *then* $A^- = B^-$.

2. *If* $\Gamma \vdash K = L : \text{kind}$ *then* $K^- = L^-$.

3. If $\Gamma, x{:}A \vdash B : K$ then $B^- = [M/x]B^-$.

4. If $\Gamma, x{:}A \vdash K : \text{kind}$ then $K^- = [M/x]K^-$.

**Proof:** By induction over the structure of the given derivations. □

We now present the algorithm in the form of three judgments.

$M \xrightarrow{\text{whr}} M'$ ($M$ *weak head reduces to* $M'$) Algorithmically, we assume $M$ is given and compute $M'$ (if $M$ is head reducible) or fail.

$\Delta \vdash M \Longleftrightarrow N : \tau$ ($M$ *is equal to* $N$ *at simple type* $\tau$) Algorithmically, we assume $\Delta$, $M$, $N$, and $\tau$ are given and we simply succeed or fail. We only apply this judgment if $M$ and $N$ have the same type $A$ and $\tau = A^-$.

$\Delta \vdash M \longleftrightarrow N : \tau$ ($M$ *is structurally equal to* $N$) Algorithmically, we assume that $\Delta$, $M$ and $N$ are given and we compute $\tau$ or fail. If successful, $\tau$ will be the approximate type of $M$ and $N$.

Note that the structural and type-directed equality are mutually recursive, while weak head reduction does not depend on the other two judgments.

**Weak Head Reduction**

$$\frac{}{(\lambda x{:}A_1.\ M_2)\ M_1 \xrightarrow{\text{whr}} [M_1/x]M_2} \qquad \frac{M_1 \xrightarrow{\text{whr}} M_1'}{M_1\ M_2 \xrightarrow{\text{whr}} M_1'\ M_2}$$

**Type-Directed Object Equality**

$$\frac{M \xrightarrow{\text{whr}} M' \quad \Delta \vdash M' \Longleftrightarrow N : \alpha}{\Delta \vdash M \Longleftrightarrow N : \alpha} \qquad \frac{N \xrightarrow{\text{whr}} N' \quad \Delta \vdash M \Longleftrightarrow N' : \alpha}{\Delta \vdash M \Longleftrightarrow N : \alpha}$$

$$\frac{\Delta \vdash M \longleftrightarrow N : \alpha}{\Delta \vdash M \Longleftrightarrow N : \alpha} \qquad \frac{\Delta, x{:}\tau_1 \vdash M\ x \Longleftrightarrow N\ x : \tau_2}{\Delta \vdash M \Longleftrightarrow N : \tau_1 \to \tau_2}$$

**Structural Object Equality**

$$\frac{x{:}\tau \text{ in } \Delta}{\Delta \vdash x \longleftrightarrow x : \tau} \qquad \frac{c{:}A \text{ in } \Sigma}{\Delta \vdash c \longleftrightarrow c : A^-}$$

$$\frac{\Delta \vdash M_1 \longleftrightarrow N_1 : \tau_2 \to \tau_1 \quad \Delta \vdash M_2 \Longleftrightarrow N_2 : \tau_2}{\Delta \vdash M_1\ M_2 \longleftrightarrow N_1\ N_2 : \tau_1}$$

We mirror these judgments at the level of families. Due to the absence of $\lambda$-abstraction at this level, the kind-directed and structural equality are rather close. However, in the later development and specifically the proof that logically related terms are algorithmically equal (Theorem 19), the distinction is still convenient.

**Kind-Directed Family Equality**

$$\frac{\Delta \vdash A \longleftrightarrow B : \text{type}^-}{\Delta \vdash A \Longleftrightarrow B : \text{type}^-} \qquad \frac{\Delta, x{:}\tau \vdash A\,x \Longleftrightarrow B\,x : \kappa}{\Delta \vdash A \Longleftrightarrow B : \tau \to \kappa}$$

$$\frac{\Delta \vdash A_1 \Longleftrightarrow B_1 : \text{type}^- \qquad \Delta, x{:}A_1^- \vdash A_2 \Longleftrightarrow B_2 : \text{type}^-}{\Delta \vdash \Pi x{:}A_1.\ A_2 \Longleftrightarrow \Pi x{:}B_1.\ B_2 : \text{type}^-}$$

**Structural Family Equality**

$$\frac{a{:}K \text{ in } \Sigma}{\Delta \vdash a \longleftrightarrow a : K^-} \qquad \frac{\Delta \vdash A \longleftrightarrow B : \tau \to \kappa \qquad \Delta \vdash M \Longleftrightarrow N : \tau}{\Delta \vdash A\,M \longleftrightarrow A\,N : \kappa}$$

**Algorithmic Kind Equality**

$$\frac{}{\Delta \vdash \text{type} \Longleftrightarrow \text{type} : \text{kind}^-} \qquad \frac{\Delta \vdash A \Longleftrightarrow B : \text{type}^- \qquad \Delta, x{:}A^- \vdash K \Longleftrightarrow L : \text{kind}^-}{\Delta \vdash \Pi x{:}A.\ K \Longleftrightarrow \Pi x{:}B.\ L : \text{kind}^-}$$

The algorithmic equality satisfies some straightforward structural properties. Weakening is required in the proof of its correctness. It does not appear that exchange, contraction, or strengthening are needed in our particular argument, but these properties can all be easily proven. Note that versions of the logical relations proofs nonetheless apply in the linear, strict, and affine λ-calculi.

**Lemma 14 (Weakening of Algorithmic Equality)**
*For each algorithmic equality judgment $J$, if $\Delta, \Delta' \vdash J$ then $\Delta, x{:}\tau, \Delta' \vdash J$.*

**Proof:** By straightforward induction over the structure of the given derivations. □

The algorithm is essentially deterministic in the sense that when comparing terms at base type we have to weakly head-normalize both sides and compare the results structurally. This is because terms that are weakly head reducible will never be considered structurally equal.

**Lemma 15 (Determinacy of Algorithmic Equality)**

1. If $M \xrightarrow{\text{whr}} M'$ and $M \xrightarrow{\text{whr}} M''$ then $M' = M''$.

2. If $\Delta \vdash M \longleftrightarrow N : \tau$ then there is no $M'$ such that $M \xrightarrow{\text{whr}} M'$.

3. If $\Delta \vdash M \longleftrightarrow N : \tau$ then there is no $N'$ such that $N \xrightarrow{\text{whr}} N'$.

4. If $\Delta \vdash M \longleftrightarrow N : \tau$ and $\Delta \vdash M \longleftrightarrow N : \tau'$ then $\tau = \tau'$.

5. If $\Delta \vdash A \longleftrightarrow B : \kappa$ and $\Delta \vdash A \longleftrightarrow B : \kappa'$ then $\kappa = \kappa'$.

**Proof:** The first part and parts (4) and (5) are immediate by structural induction. We only show the second part, since the third part is symmetric. Assume

$$\begin{array}{cc} \mathcal{S} & \mathcal{W} \\ \Delta \vdash M \longleftrightarrow N : \tau & \qquad M \xrightarrow{\text{whr}} M' \end{array} \quad \text{and}$$

for some $M'$. We now show by simultaneous induction over $\mathcal{S}$ and $\mathcal{W}$ that these assumptions are contradictory. Whenever we have constructed a judgment such there is no rule that could conclude this judgment, we say we obtain a contradiction by inversion.

**Case:**

$$\mathcal{S} = \frac{x{:}\tau \text{ in } \Delta}{\Delta \vdash x \longleftrightarrow x : \tau}$$

$x \xrightarrow{\text{whr}} M'$      Assumption ($\mathcal{W}$)
Contradiction      By inversion

**Case:** Structural equality of constants is impossible as in the case for variables.

**Case:**

$$\mathcal{S} = \frac{\begin{array}{cc}\mathcal{S}_1 & \mathcal{T}_2 \\ \Delta \vdash M_1 \longleftrightarrow N_1 : \tau_2 \to \tau_1 & \Delta \vdash M_2 \Longleftrightarrow N_2 : \tau_2\end{array}}{\Delta \vdash M_1\, M_2 \longleftrightarrow N_1\, N_2 : \tau_1}$$

Here we distinguish two subcases for the derivation $\mathcal{W}$ of $M_1\, M_2 \xrightarrow{\text{whr}} M'$.

**Subcase:**

$$\mathcal{W} = \frac{}{(\lambda x{:}A_1.\ M_1')\, M_2 \xrightarrow{\text{whr}} [M_2/x]M_1'}$$

$M_1 = (\lambda x{:}A_1.\ M_1')$      Assumption
$\Delta \vdash M_1 \longleftrightarrow N_1 : \tau_2 \to \tau_1$      Assumption ($\mathcal{S}_1$)
Contradiction      By inversion

**Subcase:**

$$\mathcal{W} = \frac{\begin{array}{c}\mathcal{W}_1 \\ M_1 \xrightarrow{\text{whr}} M_1'\end{array}}{M_1\, M_2 \xrightarrow{\text{whr}} M_1'\, M_2}$$

$\Delta \vdash M_1 \longleftrightarrow N_1 : \tau_2 \to \tau_1$      Assumption ($\mathcal{S}_1$)
Contradiction      By ind. hyp. on $\mathcal{W}_1$ and $\mathcal{S}_1$

$\square$

The completeness proof requires symmetry and transitivity of the algorithm. This would introduce some difficulty if the algorithm employed precise instead of approximate types. This is one reason why both the algorithm and later the logical relation are defined using approximate types only.

**Lemma 16 (Symmetry of Algorithmic Equality)**

1. If $\Delta \vdash M \Longleftrightarrow N : \tau$ then $\Delta \vdash N \Longleftrightarrow M : \tau$.

2. If $\Delta \vdash M \longleftrightarrow N : \tau$ then $\Delta \vdash N \longleftrightarrow M : \tau$.

3. If $\Delta \vdash A \Longleftrightarrow B : \kappa$ then $\Delta \vdash B \Longleftrightarrow A : \kappa$.

4. If $\Delta \vdash A \longleftrightarrow B : \kappa$ then $\Delta \vdash B \longleftrightarrow A : \kappa$.

5. If $\Delta \vdash K \Longleftrightarrow L : \text{kind}^-$ then $\Delta \vdash L \Longleftrightarrow K : \text{kind}^-$.

**Proof:** By simultaneous induction on the given derivations. □

**Lemma 17 (Transitivity of Algorithmic Equality)**

1. If $\Delta \vdash M \Longleftrightarrow N : \tau$ and $\Delta \vdash N \Longleftrightarrow O : \tau$ then $\Delta \vdash M \Longleftrightarrow O : \tau$.

2. If $\Delta \vdash M \longleftrightarrow N : \tau$ and $\Delta \vdash N \longleftrightarrow O : \tau$ then $\Delta \vdash M \longleftrightarrow O : \tau$.

3. If $\Delta \vdash A \Longleftrightarrow B : \kappa$ and $\Delta \vdash B \Longleftrightarrow C : \kappa$ then $\Delta \vdash A \Longleftrightarrow C : \kappa$.

4. If $\Delta \vdash A \longleftrightarrow B : \kappa$ and $\Delta \vdash B \longleftrightarrow C : \kappa$ then $\Delta \vdash A \longleftrightarrow C : \kappa$.

5. If $\Delta \vdash K \Longleftrightarrow L : \text{kind}^-$ and $\Delta \vdash L \Longleftrightarrow L' : \text{kind}^-$ then $\Delta \vdash K \Longleftrightarrow L' : \text{kind}^-$.

**Proof:** By simultaneous inductions on the structure of the given derivations. In each case, we may appeal to the induction hypothesis if one of the two derivations is strictly smaller, while the other is either smaller or the same. The proof requires determinacy (Lemma 15). We only show some cases in the proof of property (1); others are direct. Assume we are given

$$\begin{array}{cc} \mathcal{T}_L & \mathcal{T}_R \\ \Delta \vdash M \Longleftrightarrow N : \tau & \text{and} \quad \Delta \vdash N \Longleftrightarrow O : \tau \end{array}$$

We have to construct a derivation of $\Delta \vdash M \Longleftrightarrow O : \tau$. We distinguish cases for $\mathcal{T}_L$ and $\mathcal{T}_R$. In case one of them is the extensionality rule, the other must be, too, and the result follows easily from the induction hypothesis. We show the remaining cases.

**Case:**

$$\mathcal{T}_L = \dfrac{M \xrightarrow{\text{whr}} M' \qquad \overset{\mathcal{T}'_L}{\Delta \vdash M' \Longleftrightarrow N : \alpha}}{\Delta \vdash M \Longleftrightarrow N : \alpha}$$

where $\mathcal{T}_R$ is arbitrary.

$\Delta \vdash M' \Longleftrightarrow O : \alpha$                By ind. hyp. (1) on $\mathcal{T}'_L$ and $\mathcal{T}_R$
$\Delta \vdash M \Longleftrightarrow O : \alpha$                    By rule (whr left)

**Case:**

$$\mathcal{T}_R = \dfrac{O \xrightarrow{\text{whr}} O' \qquad \overset{\mathcal{T}'_R}{\Delta \vdash N \Longleftrightarrow O' : \alpha}}{\Delta \vdash N \Longleftrightarrow O : \alpha}$$

where $\mathcal{T}_L$ arbitrary.

$$\Delta \vdash M \Longleftrightarrow O' : \alpha \qquad \text{By ind. hyp. (1) on } \mathcal{T}_L \text{ and } \mathcal{T}'_R$$
$$\Delta \vdash M \Longleftrightarrow O : \alpha \qquad \text{By rule (whr right)}$$

**Case:**

$$\mathcal{T}_L = \dfrac{N \xrightarrow{\text{whr}} N' \quad \Delta \vdash M \stackrel{\mathcal{T}'_L}{\Longleftrightarrow} N' : \alpha}{\Delta \vdash M \Longleftrightarrow N : \alpha} \quad \text{and} \quad \mathcal{T}_R = \dfrac{N \xrightarrow{\text{whr}} N'' \quad \Delta \vdash N'' \stackrel{\mathcal{T}'_R}{\Longleftrightarrow} O : \alpha}{\Delta \vdash N \Longleftrightarrow O : \alpha}$$

$N' = N''$   By determinacy of weak head reduction (Lemma 15(1))
$\Delta \vdash M \Longleftrightarrow O : \alpha$   By ind. hyp. (1) on $\mathcal{T}'_L$ and $\mathcal{T}'_R$.

**Case:**

$$\mathcal{T}_L = \dfrac{N \xrightarrow{\text{whr}} N' \quad \Delta \vdash M \stackrel{\mathcal{T}'_L}{\Longleftrightarrow} N' : \alpha}{\Delta \vdash M \Longleftrightarrow N : \alpha} \quad \text{and} \quad \mathcal{T}_R = \dfrac{\Delta \vdash N \stackrel{\mathcal{S}_R}{\longleftrightarrow} O : \alpha}{\Delta \vdash N \Longleftrightarrow O : \alpha}$$

This case is impossible by determinacy of algorithmic equality (Lemma 15(2)).

**Case:**

$$\mathcal{T}_L = \dfrac{\Delta \vdash M \stackrel{\mathcal{S}_L}{\longleftrightarrow} N : \alpha}{\Delta \vdash M \Longleftrightarrow N : \alpha} \quad \text{and} \quad \mathcal{T}_R = \dfrac{N \xrightarrow{\text{whr}} N' \quad \Delta \vdash N' \stackrel{\mathcal{T}'_R}{\Longleftrightarrow} O : \alpha}{\Delta \vdash N \Longleftrightarrow O : \alpha}$$

This case is impossible by determinacy of algorithmic equality (Lemma 15(3)).

**Case:**

$$\mathcal{T}_L = \dfrac{\Delta \vdash M \stackrel{\mathcal{S}_L}{\longleftrightarrow} N : \alpha}{\Delta \vdash M \Longleftrightarrow N : \alpha} \quad \text{and} \quad \mathcal{T}_R = \dfrac{\Delta \vdash N \stackrel{\mathcal{S}_R}{\longleftrightarrow} O : \alpha}{\Delta \vdash N \Longleftrightarrow O : \alpha}$$

$\Delta \vdash M \longleftrightarrow O : \alpha$   By ind. hyp. (2) on $\mathcal{S}_L$ and $\mathcal{S}_R$
$\Delta \vdash M \Longleftrightarrow O : \alpha$   By rule

□

## 4 Completeness of Algorithmic Equality

In this section we develop the completeness theorem for the type-directed equality algorithm. That is, if two terms are definitionally equal, the algorithm will succeed. The goal is to present a flexible and modular technique which can be adapted easily to related type theories, such as the one underlying the linear logical framework [CP98, VC01], one based on ordered linear logic [PP99, Pol01], or one including subtyping [Pfe93] or proof irrelevance and intensional types [Pfe01]. Other techniques presented in the literature, particularly those based on a notion of $\eta$-reduction, do not seem to adapt well to these richer theories.

The central idea is to proceed by an argument via logical relations defined inductively on the approximate type of an object, where the approximate type arises from erasing all dependencies in an LF type.

The completeness direction of the correctness proof for type-directed equality states:

If $\Gamma \vdash M = N : A$ then $\Gamma^- \vdash M \Longleftrightarrow N : A^-$.

One would like to prove this by induction on the structure of the derivation for the given equality. However, such a proof attempt fails at the case for application. Instead we define a logical relation $\Delta \vdash M = N \in [\![\tau]\!]$ that provides a stronger induction hypothesis so that both

1. if $\Gamma \vdash M = N : A$ then $\Gamma^- \vdash M = N \in [\![A^-]\!]$, and
2. if $\Gamma^- \vdash M = N \in [\![A^-]\!]$ then $\Gamma^- \vdash M \Longleftrightarrow N \in A^-$,

can be proved.

## 4.1 A Kripke Logical Relation

We define a Kripke logical relation inductively on simple types. At base type we require the property we eventually would like to prove. At higher types we reduce the property to those for simpler types. We also extend it further to include substitutions, where it is defined by induction over the structure of the matching context.

We say that a context $\Delta'$ extends $\Delta$ (written $\Delta' \geq \Delta$) if $\Delta'$ contains all declarations in $\Delta$ and possibly more.

1. $\Delta \vdash M = N \in [\![\alpha]\!]$ iff $\Delta \vdash M \Longleftrightarrow N : \alpha$.

2. $\Delta \vdash M = N \in [\![\tau_1 \to \tau_2]\!]$ iff for every $\Delta'$ extending $\Delta$ and for all $M_1$ and $N_1$ such that $\Delta' \vdash M_1 = N_1 \in [\![\tau_1]\!]$ we have $\Delta' \vdash M\,M_1 = N\,N_1 \in [\![\tau_2]\!]$.

3. $\Delta \vdash A = B \in [\![\text{type}^-]\!]$ iff $\Delta \vdash A \Longleftrightarrow B : \text{type}^-$.

4. $\Delta \vdash A = B \in [\![\tau \to \kappa]\!]$ iff for every $\Delta'$ extending $\Delta$ and for all $M$ and $N$ such that $\Delta' \vdash M = N \in [\![\tau]\!]$ we have $\Delta' \vdash A\,M = B\,N \in [\![\kappa]\!]$.

5. $\Delta \vdash \sigma = \theta \in [\![\cdot]\!]$ iff $\sigma = \cdot$ and $\theta = \cdot$.

6. $\Delta \vdash \sigma = \theta \in [\![\Theta, x{:}\tau]\!]$ iff $\sigma = (\sigma', M/x)$ and $\theta = (\theta', N/x)$ where $\Delta \vdash \sigma' = \theta' \in [\![\Theta]\!]$ and $\Delta \vdash M = N \in [\![\tau]\!]$.

Four general structural properties of the logical relations that we can show directly by induction are exchange, weakening, contraction, and strengthening. We will use only weakening.

**Lemma 18 (Weakening of the Logical Relations)** *For all logical relations $R$, if $\Delta, \Delta' \vdash R$ then $\Delta, x{:}\tau, \Delta' \vdash R$.*

**Proof:** By induction on the structure of the definition of $R$ (either simple type, kind, or context). We show only the proof for the relation on types: If $\Delta, \Delta' \vdash M \in [\![\tau]\!]$ then $\Delta, x{:}\theta, \Delta' \vdash M = N \in [\![\tau]\!]$.

**Case:** $\tau = \alpha$.

| | |
|---|---:|
| $\Delta, \Delta' \vdash M = N \in [\![\alpha]\!]$ | Assumption |
| $\Delta, \Delta' \vdash M \Longleftrightarrow N : \alpha$ | By definition of $[\![\alpha]\!]$ |
| $\Delta, x{:}\theta, \Delta' \vdash M \Longleftrightarrow N : \alpha$ | By weakening (Lemma 14) |
| $\Delta, x{:}\theta, \Delta' \vdash M = N \in [\![\alpha]\!]$ | By definition of $[\![\alpha]\!]$ |

**Case:** $\tau = \tau_1 \to \tau_2$.

$$\begin{array}{ll}
\Delta, \Delta' \vdash M = N \in [\![\tau_1 \to \tau_2]\!] & \text{Assumption} \\
\Delta_+, x{:}\theta, \Delta'_+ \vdash M_1 = N_1 \in [\![\tau_1]\!] & \\
\quad \text{for arbitrary } \Delta_+ \geq \Delta \text{ and } \Delta'_+ \geq \Delta' & \text{New assumption} \\
(\Delta_+, x{:}\theta, \Delta'_+) \geq (\Delta, \Delta') & \text{By definition of } \geq \\
\Delta_+, x{:}\theta, \Delta'_+ \vdash M\,M_1 = N\,N_1 \in [\![\tau_2]\!] & \text{By definition of } [\![\tau_1 \to \tau_2]\!] \text{ and assumption} \\
\Delta, x{:}\theta, \Delta' \vdash M = N \in [\![\tau_1 \to \tau_2]\!] & \text{By definition of } [\![\tau_1 \to \tau_2]\!]
\end{array}$$

□

## 4.2 Logically Related Terms are Algorithmically Equal

It is straightforward to show that logically related terms are considered identical by the algorithm. This proof always proceeds by induction on the structure of the type. A small insight may be required to arrive at the necessary generalization of the induction hypothesis. Here, this involves the statement that structurally equal terms are logically related. This has an important consequence we will need later on, namely that variables and constants are logically related to themselves.

**Theorem 19 (Logically Related Terms are Algorithmically Equal)**

1. *If* $\Delta \vdash M = N \in [\![\tau]\!]$ *then* $\Delta \vdash M \iff N : \tau$.

2. *If* $\Delta \vdash A = B \in [\![\kappa]\!]$, *then* $\Delta \vdash A \iff B : \kappa$.

3. *If* $\Delta \vdash M \longleftrightarrow N : \tau$ *then* $\Delta \vdash M = N \in [\![\tau]\!]$.

4. *If* $\Delta \vdash A \longleftrightarrow B : \kappa$ *then* $\Delta \vdash A = B \in [\![\kappa]\!]$.

**Proof:** By simultaneous induction on the structure of $\tau$.

**Case:** $\tau = \alpha$, part 1.

$$\begin{array}{ll}
\Delta \vdash M = N \in [\![\alpha]\!] & \text{Assumption} \\
\Delta \vdash M \iff N : \alpha & \text{By definition of } [\![\alpha]\!]
\end{array}$$

**Case:** $\kappa = \text{type}^-$, part 2.

$$\begin{array}{ll}
\Delta \vdash A = B \in [\![\text{type}^-]\!] & \text{Assumption} \\
\Delta \vdash A \iff B : \text{type}^- & \text{By definition of } [\![\text{type}^-]\!]
\end{array}$$

**Case:** $\tau = \alpha$, part 3.

$$\begin{array}{ll}
\Delta \vdash M \longleftrightarrow N : \alpha & \text{Assumption} \\
\Delta \vdash M \iff N : \alpha & \text{By rule} \\
\Delta \vdash M = N \in [\![\alpha]\!] & \text{By definition of } [\![\alpha]\!]
\end{array}$$

**Case:** $\kappa = \text{type}^-$, part 4.

$$\Delta \vdash A \longleftrightarrow B : \text{type}^- \qquad \text{Assumption}$$
$$\Delta \vdash A \Longleftrightarrow B : \text{type}^- \qquad \text{By rule}$$
$$\Delta \vdash A = B \in [\![\text{type}^-]\!] \qquad \text{By definition of } [\![\text{type}^-]\!]$$

**Case:** $\tau = \tau_1 \to \tau_2$, part 1.

$$\Delta \vdash M = N \in [\![\tau_1 \to \tau_2]\!] \qquad \text{Assumption}$$
$$\Delta, x{:}\tau_1 \vdash x \longleftrightarrow x : \tau_1 \qquad \text{By rule}$$
$$\Delta, x{:}\tau_1 \vdash x = x \in [\![\tau_1]\!] \qquad \text{By i.h. 3 on } \tau_1$$
$$\Delta, x{:}\tau_1 \vdash M\,x = N\,x \in [\![\tau_2]\!] \qquad \text{By definition of } [\![\tau_1 \to \tau_2]\!]$$
$$\Delta, x{:}\tau_1 \vdash M\,x \Longleftrightarrow N\,x : \tau_2 \qquad \text{By i.h. 1 on } \tau_2$$
$$\Delta \vdash M \Longleftrightarrow N : \tau_1 \to \tau_2 \qquad \text{By rule}$$

**Case:** $\kappa = \tau_1 \to \kappa_2$, part 2.

$$\Delta \vdash A = B \in [\![\tau_1 \to \kappa_2]\!] \qquad \text{Assumption}$$
$$\Delta, x{:}\tau_1 \vdash x \longleftrightarrow x : \tau_1 \qquad \text{By rule}$$
$$\Delta, x{:}\tau_1 \vdash x = x \in [\![\tau_1]\!] \qquad \text{By i.h. 3 on } \tau_1$$
$$\Delta, x{:}\tau_1 \vdash A\,x = B\,x \in [\![\kappa_2]\!] \qquad \text{By definition of } [\![\tau_1 \to \kappa_2]\!]$$
$$\Delta, x{:}\tau_1 \vdash A\,x \Longleftrightarrow B\,x : \kappa_2 \qquad \text{By i.h. 2 on } \kappa_2$$
$$\Delta \vdash A \Longleftrightarrow B : \tau_1 \to \kappa_2 \qquad \text{By rule}$$

**Case:** $\tau = \tau_1 \to \tau_2$, part 3.

$$\Delta \vdash M \longleftrightarrow N : \tau_1 \to \tau_2 \qquad \text{Assumption}$$
$$\Delta_+ \vdash M_1 = N_1 \in [\![\tau_1]\!] \text{ for an arbitrary } \Delta_+ \geq \Delta \qquad \text{New assumption}$$
$$\Delta_+ \vdash M_1 \Longleftrightarrow N_1 : \tau_1 \qquad \text{By i.h. 1 on } \tau_1$$
$$\Delta_+ \vdash M \longleftrightarrow N : \tau_1 \to \tau_2 \qquad \text{By weakening (Lemma 14)}$$
$$\Delta_+ \vdash M\,M_1 \longleftrightarrow N\,N_1 : \tau_2 \qquad \text{By rule}$$
$$\Delta_+ \vdash M\,M_1 = N\,N_1 \in [\![\tau_2]\!] \qquad \text{By i.h. 3 on } \tau_2$$
$$\Delta \vdash M = N \in [\![\tau_1 \to \tau_2]\!] \qquad \text{By definition of } [\![\tau_1 \to \tau_2]\!]$$

**Case:** $\kappa = \tau_1 \to \kappa_2$, part 4.

$$\Delta \vdash A \longleftrightarrow B : \tau_1 \to \kappa_2 \qquad \text{Assumption}$$
$$\Delta_+ \vdash M_1 = N_1 \in [\![\tau_1]\!] \text{ for an arbitrary } \Delta_+ \geq \Delta \qquad \text{New assumption}$$
$$\Delta_+ \vdash M_1 \Longleftrightarrow N_1 : \tau_1 \qquad \text{By i.h. 1 on } \tau_1$$
$$\Delta_+ \vdash A \longleftrightarrow B : \tau_1 \to \kappa_2 \qquad \text{By weakening (Lemma 14)}$$
$$\Delta_+ \vdash A\,M_1 \longleftrightarrow B\,N_1 \in \kappa_2 \qquad \text{By rule}$$
$$\Delta_+ \vdash A\,M_1 = B\,N_1 \in [\![\kappa_2]\!] \qquad \text{By i.h. 4 on } \kappa_2$$
$$\Delta \vdash A = B \in [\![\tau_1 \to \kappa_2]\!] \qquad \text{By definition of } [\![\tau_1 \to \kappa_2]\!]$$

□

## 4.3 Definitionally Equal Terms are Logically Related

The other part of the logical relations argument states that two equal terms are logically related. This requires a sequence of lemmas regarding algorithmic equality and the logical relation.

**Lemma 20 (Closure under Head Expansion)**

1. If $M \xrightarrow{\text{whr}} M'$ and $\Delta \vdash M' = N \in [\![\tau]\!]$ then $\Delta \vdash M = N \in [\![\tau]\!]$.

2. If $N \xrightarrow{\text{whr}} N'$ and $\Delta \vdash M = N' \in [\![\tau]\!]$ then $\Delta \vdash M = N \in [\![\tau]\!]$.

**Proof:** Each part follows by induction on the structure of $\tau$. We show only the first, since the second is symmetric.

**Case:** $\tau = \alpha$.

| | |
|---|---:|
| $M \xrightarrow{\text{whr}} M'$ | Assumption |
| $\Delta \vdash M' = N \in [\![\alpha]\!]$ | Assumption |
| $\Delta \vdash M' \Longleftrightarrow N : \alpha$ | By definition of $[\![\alpha]\!]$ |
| $\Delta \vdash M \Longleftrightarrow N : \alpha$ | By rule (whr) |
| $\Delta \vdash M = N \in [\![\alpha]\!]$ | By definition of $[\![\alpha]\!]$ |

**Case:** $\tau = \tau_1 \to \tau_2$.

| | |
|---|---:|
| $M \xrightarrow{\text{whr}} M'$ | Assumption |
| $\Delta \vdash M' = N \in [\![\tau_1 \to \tau_2]\!]$ | Assumption |
| $\Delta_+ \vdash M_1 = N_1 \in [\![\tau_1]\!]$ for $\Delta_+ \geq \Delta$ | New assumption |
| $\Delta_+ \vdash M' \, M_1 = N \, N_1 \in [\![\tau_2]\!]$ | By definition of $[\![\tau_1 \to \tau_2]\!]$ |
| $M \, M_1 \xrightarrow{\text{whr}} M' \, M_1$ | By rule |
| $\Delta_+ \vdash M \, M_1 = N \, N_1 \in [\![\tau_2]\!]$ | By i.h. on $\tau_2$ |
| $\Delta \vdash M = N \in [\![\tau_1 \to \tau_2]\!]$ | By definition of $[\![\tau_1 \to \tau_2]\!]$ |

□

**Lemma 21 (Symmetry of the Logical Relations)**

1. If $\Delta \vdash M = N \in [\![\tau]\!]$ then $\Delta \vdash N = M \in [\![\tau]\!]$.

2. If $\Delta \vdash A = B \in [\![\kappa]\!]$ then $\Delta \vdash B = A \in [\![\kappa]\!]$.

3. If $\Delta \vdash \sigma = \theta \in [\![\Theta]\!]$ then $\Delta \vdash \theta = \sigma \in [\![\Theta]\!]$.

**Proof:** By induction on the structure of $\tau$, $\kappa$, and $\Theta$, using Lemma 16. We show some representative cases.

**Case:** $\tau = \alpha$.

| | |
|---|---:|
| $\Delta \vdash M = N \in [\![\alpha]\!]$ | Assumption |
| $\Delta \vdash M \Longleftrightarrow N : \alpha$ | By definition of $[\![\alpha]\!]$ |
| $\Delta \vdash N \Longleftrightarrow M : \alpha$ | By symmetry of type-directed equality (Lemma 16) |
| $\Delta \vdash N = M \in [\![\alpha]\!]$ | By definition of $[\![\alpha]\!]$ |

**Case:** $\tau = \tau_1 \to \tau_2$.

| | |
|---|---:|
| $\Delta \vdash M = N \in [\![\tau_1 \to \tau_2]\!]$ | Assumption |
| $\Delta_+ \vdash N_1 = M_1 \in [\![\tau_1]\!]$ for $\Delta_+ \geq \Delta$ | New assumption |
| $\Delta_+ \vdash M_1 = N_1 \in [\![\tau_1]\!]$ | By i.h. on $\tau_1$ |
| $\Delta_+ \vdash M\,M_1 = N\,N_1 \in [\![\tau_2]\!]$ | By definition of $[\![\tau_1 \to \tau_2]\!]$ |
| $\Delta_+ \vdash N\,N_1 = M\,M_1 \in [\![\tau_2]\!]$ | By i.h. on $\tau_2$ |
| $\Delta \vdash N = M \in [\![\tau_1]\!]$ | By definition of $[\![\tau_1 \to \tau_2]\!]$ |

□

**Lemma 22 (Transitivity of the Logical Relations)**

1. If $\Delta \vdash M = N \in [\![\tau]\!]$ and $\Delta \vdash N = O \in [\![\tau]\!]$ then $\Delta \vdash M = O \in [\![\tau]\!]$.

2. If $\Delta \vdash A = B \in [\![\kappa]\!]$ and $\Delta \vdash B = C \in [\![\kappa]\!]$ then $\Delta \vdash A = C \in [\![\kappa]\!]$.

3. If $\Delta \vdash \sigma = \theta \in [\![\Theta]\!]$ and $\Delta \vdash \theta = \delta \in [\![\Theta]\!]$ then $\Delta \vdash \sigma = \delta \in [\![\Theta]\!]$.

**Proof:** By induction on the structure of $\tau$, $\kappa$, and $\Theta$, using Lemma 17. We show some representative cases.

**Case:** $\tau = \alpha$. Then the properties follows from the definition of $[\![\alpha]\!]$ and the transitivity of type-directed equality (Lemma 17).

| | |
|---|---:|
| $\Delta \vdash M = N \in [\![\alpha]\!]$ | Assumption |
| $\Delta \vdash N = O \in [\![\alpha]\!]$ | Assumption |
| $\Delta \vdash M \Longleftrightarrow N : \alpha$ | By definition of $[\![\alpha]\!]$ |
| $\Delta \vdash N \Longleftrightarrow O : \alpha$ | By definition of $[\![\alpha]\!]$ |
| $\Delta \vdash M \Longleftrightarrow O : \alpha$ | By transitivity of type-directed equality (Lemma 17) |
| $\Delta \vdash M = O \in [\![\alpha]\!]$ | By definition of $[\![\alpha]\!]$ |

**Case:** $\tau = \tau_1 \to \tau_2$.

| | |
|---|---:|
| $\Delta \vdash M = N \in [\![\tau_1 \to \tau_2]\!]$ | Assumption |
| $\Delta \vdash N = O \in [\![\tau_1 \to \tau_2]\!]$ | Assumption |
| $\Delta_+ \vdash M_1 = O_1 \in [\![\tau_1]\!]$ for $\Delta_+ \geq \Delta$ | New assumption |
| $\Delta_+ \vdash M\,M_1 = N\,O_1 \in [\![\tau_2]\!]$ | By definition of $[\![\tau_1 \to \tau_2]\!]$ |
| $\Delta_+ \vdash O_1 = M_1 \in [\![\tau_1]\!]$ | By symmetry (Lemma 21) |
| $\Delta_+ \vdash O_1 = O_1 \in [\![\tau_1]\!]$ | By i.h. on $\tau_1$ |
| $\Delta_+ \vdash N\,O_1 = O\,O_1 \in [\![\tau_2]\!]$ | By definition of $[\![\tau_1 \to \tau_2]\!]$ |
| $\Delta_+ \vdash M\,M_1 = O\,O_1 \in [\![\tau_2]\!]$ | By i.h. on $\tau_2$ |
| $\Delta \vdash M = O \in [\![\tau_1 \to \tau_2]\!]$ | By definition of $[\![\tau_1 \to \tau_2]\!]$ |

□

**Lemma 23 (Definitionally Equal Terms are Logically Related under Substitutions)**

1. If $\Gamma \vdash M = N : A$ and $\Delta \vdash \sigma = \theta \in [\![\Gamma^-]\!]$ then $\Delta \vdash M[\sigma] = N[\theta] \in [\![A^-]\!]$.

2. If $\Gamma \vdash A = B : K$ and $\Delta \vdash \sigma = \theta \in [\![\Gamma^-]\!]$ then $\Delta \vdash A[\sigma] = B[\theta] \in [\![K^-]\!]$.

**Proof:** By induction on the derivation $\mathcal{D}$ of definitional equality, using the prior lemmas in this section. For this argument, some subderivations of the equality judgment are unnecessary (in particular, those establishing the validity of certain types). We elide those premises and write "..." instead.

**Case:**

$$\mathcal{D} = \frac{x{:}A \text{ in } \Gamma}{\Gamma \vdash x = x : A}$$

| | |
|---|---:|
| $\Delta \vdash \sigma = \theta \in [\![\Gamma^-]\!]$ | |
| $\Delta \vdash M = N \in [\![A^-]\!]$ for $M/x$ in $\sigma$ and $N/x$ in $\theta$ | By definition of $[\![\Gamma^-]\!]$ |
| $\Delta \vdash x[\sigma] = x[\theta] \in [\![A^-]\!]$ | By definition of substitution |

**Case:**

$$\mathcal{D} = \frac{c{:}A \text{ in } \Sigma}{\Gamma \vdash c = c : A}$$

| | |
|---|---:|
| $\Delta \vdash c \longleftrightarrow c \in [\![A^-]\!]$ | By rule |
| $\Delta \vdash c = c \in [\![A^-]\!]$ | By Theorem 19(3) |
| $\Delta \vdash c[\sigma] = c[\theta] \in [\![A^-]\!]$ | By definition of substitution |

**Case:**

$$\mathcal{D} = \frac{\begin{array}{c}\mathcal{D}_1 \\ \Gamma \vdash M_1 = N_1 : \Pi x{:}A_2.\ A_1\end{array} \quad \begin{array}{c}\mathcal{D}_2 \\ \Gamma \vdash M_2 = N_2 : A_2\end{array}}{\Gamma \vdash M_1\ M_2 = N_1\ N_2 : [M_2/x]A_1}$$

| | |
|---|---:|
| $\Delta \vdash M_1[\sigma] = N_1[\theta] \in [\![A_2^- \to A_1^-]\!]$ | By i.h. on $\mathcal{D}_1$ |
| $\Delta \vdash M_2[\sigma] = N_2[\theta] \in [\![A_2^-]\!]$ | By i.h. on $\mathcal{D}_2$ |
| $\Delta \vdash (M_1[\sigma])(M_2[\sigma]) = (N_1[\theta])(N_2[\theta]) \in [\![A_1^-]\!]$ | By definition of $[\![\tau_2 \to \tau_1]\!]$ |
| $\Delta \vdash (M_1\ M_2)[\sigma] = (N_1\ N_2)[\theta] \in [\![A_1^-]\!]$ | By definition of substitution |

**Case:**

$$\mathcal{D} = \frac{\cdots \quad \begin{array}{c}\mathcal{D}_2 \\ \Gamma, x{:}A_1 \vdash M_2 = N_2 : A_2\end{array}}{\Gamma \vdash \lambda x{:}A_1'.\ M_2 = \lambda x{:}A_1''.\ N_2 : \Pi x{:}A_1.\ A_2}$$

| | |
|---|---:|
| $\Delta_+ \vdash M_1 = N_1 \in [\![A_1^-]\!]$ for $\Delta_+ \geq \Delta$ | New assumption |
| $\Delta_+ \vdash \sigma = \theta \in [\![\Gamma^-]\!]$ | By weakening (Lemma 18) |
| $\Delta_+ \vdash (\sigma, M_1/x) = (\theta, N_1/x) \in [\![\Gamma^-, x{:}A_1^-]\!]$ | By definition of $[\![\Delta, x{:}\tau]\!]$ |
| $\Delta_+ \vdash M_2[\sigma, M_1/x] = N_2[\theta, N_1/x] \in [\![A_2^-]\!]$ | By i.h. on $\mathcal{D}_2$ |
| $\Delta_+ \vdash (\lambda x{:}A_1'.\ M_2[\sigma, x/x])\ M_1 = N_2[\theta, N_1/x] \in [\![A_2^-]\!]$ | |

$$\begin{aligned}
&\Delta_+ \vdash (\lambda x{:}A'_1.\ M_2[\sigma, x/x])\ M_1 = (\lambda x{:}A''_1.\ N_2[\theta, x/x])\ N_1 \in [\![A_2^-]\!] && \text{By closure under head expansion (Lemma 20)}\\
&\Delta_+ \vdash ((\lambda x{:}A'_1.\ M_2)[\sigma])\ M_1 = ((\lambda x{:}A''_1.\ N_2)[\theta])\ N_1 \in [\![A_2^-]\!] && \text{By closure under head expansion (Lemma 20)}\\
&&& \text{By properties of substitution}\\
&\Delta \vdash (\lambda x{:}A'_1.\ M_2)[\sigma] = (\lambda x{:}A''_1.\ N_2)[\theta] \in [\![A_1^- \to A_2^-]\!] && \text{By definition of } [\![\tau_1 \to \tau_2]\!]
\end{aligned}$$

**Case:**

$$\mathcal{D} = \quad \cdots \quad \dfrac{\begin{array}{c}\mathcal{D}_2\\ \Gamma, x{:}A_1 \vdash M\ x = N\ x : A_2\end{array}}{\Gamma \vdash M = N : \Pi x{:}A_1.\ A_2}$$

$$\begin{aligned}
&\Delta_+ \vdash M_1 = N_1 \in [\![A_1^-]\!] \text{ for } \Delta_+ \geq \Delta && \text{New assumption}\\
&\Delta_+ \vdash \sigma = \theta \in [\![\Gamma^-]\!] && \text{By weakening (Lemma 18)}\\
&\Delta_+ \vdash (\sigma, M_1/x) = (\theta, N_1/x) \in [\![\Gamma^-, x{:}A_1^-]\!] && \text{By definition of } [\![\Delta, x{:}\tau]\!]\\
&\Delta_+ \vdash (M\ x)[\sigma, M_1/x] = (N\ x)[\theta, N_1/x] \in [\![A_2^-]\!] && \text{By i.h. on } \mathcal{D}_2\\
&\Delta_+ \vdash M[\sigma]\ M_1 = N[\theta]\ N_1 \in [\![A_2^-]\!] && \text{By properties of substitution}\\
&\Delta \vdash M[\sigma] = N[\theta] \in [\![A_1^- \to A_2^-]\!] && \text{By definition of } [\![\tau_1 \to \tau_2]\!]
\end{aligned}$$

**Case:**

$$\mathcal{D} = \quad \cdots \quad \dfrac{\begin{array}{cc}\mathcal{D}_2 & \mathcal{D}_1\\ \Gamma, x{:}A_1 \vdash M_2 = N_2 : A_2 & \Gamma \vdash M_1 = N_1 : A_1\end{array}}{\Gamma \vdash (\lambda x{:}A_1.\ M_2)\ M_1 = [N_1/x]N_2 : [M_1/x]A_2}$$

$$\begin{aligned}
&\Delta \vdash \sigma = \theta \in [\![\Gamma^-]\!] && \text{Assumption}\\
&\Delta \vdash M_1[\sigma] = N_1[\theta] \in [\![A_1^-]\!] && \text{By i.h. on } \mathcal{D}_1\\
&\Delta \vdash (\sigma, M_1[\sigma]/x) = (\theta, N_1[\theta]/x) \in [\![\Gamma^-, x{:}A_1^-]\!] && \text{By definition of } [\![\Theta, x{:}\tau_1]\!]\\
&\Delta \vdash M_2[\sigma, M_1[\sigma]/x] = N_2[\theta, N_1[\theta]/x] \in [\![A_2^-]\!] && \text{By i.h. on } \mathcal{D}_2\\
&\Delta \vdash [M_1[\sigma]/x](M_2[\sigma, x/x]) = N_2[\theta, N_1[\theta]/x] \in [\![A_2^-]\!] && \text{By properties of substitution}\\
&\Delta \vdash (\lambda x{:}A_1.\ M_2[\sigma, x/x])(M_1[\sigma]) = N_1[\theta, N_1[\theta]/x] \in [\![A_2^-]\!] &&\\
&&& \text{By closure under head expansion (Lemma 20)}\\
&\Delta \vdash ((\lambda x{:}A_1.\ M_2)\ M_1)[\sigma] = ([N_1/x]N_2)[\theta] \in [\![A_2^-]\!] && \text{By properties of substitution}\\
&\Delta \vdash ((\lambda x{:}A_1.\ M_2)\ M_1)[\sigma] = ([N_1/x]N_2)[\theta] \in [\![[M_1/x]A_2^-]\!] &&\\
&&& \text{By erasure preservation (Lemma 13)}
\end{aligned}$$

**Case:**

$$\mathcal{D} = \dfrac{\begin{array}{c}\mathcal{D}'\\ \Gamma \vdash N = M : A\end{array}}{\Gamma \vdash M = N : A}$$

$$\begin{aligned}
&\Delta \vdash \sigma = \theta \in [\![\Gamma^-]\!] && \text{Assumption}\\
&\Delta \vdash \theta = \sigma \in [\![\Gamma^-]\!] && \text{By symmetry (Lemma 21)}\\
&\Delta \vdash N[\theta] = M[\sigma] \in [\![A^-]\!] && \text{By i.h. on } \mathcal{D}'\\
&\Delta \vdash M[\sigma] = N[\theta] \in [\![A^-]\!] && \text{By symmetry (Lemma 21)}
\end{aligned}$$

**Case:**

$$\mathcal{D} = \dfrac{\begin{array}{cc}\mathcal{D}_1 & \mathcal{D}_2 \\ \Gamma \vdash M = O : A & \Gamma \vdash O = N : A\end{array}}{\Gamma \vdash M = N : A}$$

| | |
|---|---:|
| $\Delta \vdash \sigma = \theta \in [\![\Gamma^-]\!]$ | Assumption |
| $\Delta \vdash \theta = \sigma \in [\![\Gamma^-]\!]$ | By symmetry (Lemma 21) |
| $\Delta \vdash \theta = \theta \in [\![\Gamma^-]\!]$ | By transitivity (Lemma 22) |
| $\Delta \vdash M[\sigma] = O[\theta] \in [\![A^-]\!]$ | By i.h. on $\mathcal{D}_1$ |
| $\Delta \vdash O[\theta] = N[\theta] \in [\![A^-]\!]$ | By i.h. on $\mathcal{D}_2$ |
| $\Delta \vdash M[\sigma] = N[\theta] \in [\![A^-]\!]$ | By transitivity (Lemma 22) |

**Case:**

$$\mathcal{D} = \dfrac{\begin{array}{cc}\mathcal{D}_1 & \\ \Gamma \vdash M = N : B & \Gamma \vdash B = A : \text{type}\end{array}}{\Gamma \vdash M = N : A}$$

| | |
|---|---:|
| $\Delta \vdash M[\sigma] = N[\theta] \in B^-$ | By i.h. on $\mathcal{D}_1$ |
| $\Delta \vdash M[\sigma] = N[\theta] \in A^-$ | By erasure preservation (Lemma 13) |

**Case:** $\Gamma \vdash a = a : K$. As for constants $c$.

**Case:** $\Gamma \vdash A_1\, M_2 = B_1\, N_2 : [M_2/x]K_1$. As for applications $M_1\, M_2$.

**Case:**

$$\mathcal{D} = \dfrac{\begin{array}{cc}\mathcal{D}_1 & \mathcal{D}_2 \\ \Gamma \vdash A_1 = B_1 : \text{type} & \Gamma, x{:}A_1 \vdash A_2 = B_2 : \text{type}\end{array}}{\Gamma \vdash \Pi x{:}A_1.\, A_2 = \Pi x{:}B_1.\, B_2 : \text{type}}$$

| | |
|---|---:|
| $\Delta \vdash A_1[\sigma] = B_1[\theta] \in [\![\text{type}^-]\!]$ | By i.h. on $\mathcal{D}_1$ |
| $\Delta \vdash A_1[\sigma] \longleftrightarrow B_1[\theta] : \text{type}^-$ | By definition of $[\![\text{type}^-]\!]$ |
| $\Delta, x{:}A_1^- \vdash x \longleftrightarrow x : A_1^-$ | By rule |
| $\Delta, x{:}A_1^- \vdash x = x \in [\![A_1^-]\!]$ | By Theorem 19(3) |
| $\Delta, x{:}A_1^- \vdash (\sigma, x/x) = (\theta, x/x) \in [\![\Gamma^-, x{:}A_1^-]\!]$ | By definition of $[\![\Theta, x{:}\tau_1]\!]$ |
| $\Delta, x{:}A_1^- \vdash A_2[\sigma, x/x] = B_2[\theta, x/x] \in [\![\text{type}^-]\!]$ | By i.h. on $\mathcal{D}_2$ |
| $\Delta, x{:}A_1^- \vdash A_2[\sigma, x/x] \Longleftrightarrow B_2[\theta, x/x] : \text{type}^-$ | By definition of $[\![\text{type}^-]\!]$ |
| $\Delta \vdash \Pi x{:}A_1[\sigma].\, A_2[\sigma, x/x] \Longleftrightarrow \Pi x{:}B_1[\theta].\, B_2[\theta, x/x] : \text{type}^-$ | By rule |
| $\Delta \vdash \Pi x{:}A_1[\sigma].\, A_2[\sigma, x/x] = \Pi x{:}B_1[\theta].\, B_2[\theta, x/x] \in [\![\text{type}^-]\!]$ | By definition of $[\![\text{type}^-]\!]$ |
| $\Delta \vdash (\Pi x{:}A_1.\, A_2)[\sigma] = (\Pi x{:}B_1.\, B_2)[\theta] \in [\![\text{type}^-]\!]$ | By definition of substitution |

**Case:** Family symmetry rule. As for the object-level symmetry.

**Case:** Family transitivity rule. As for the object-level transitivity.

**Case:** Kind conversion rule. As for type conversion rule.

**Lemma 24 (Identity Substitutions are Logically Related)**
$\Gamma^- \vdash \text{id}_\Gamma = \text{id}_\Gamma \in [\![\Gamma^-]\!]$.

**Proof:** By definition of $[\![\Gamma^-]\!]$ and part (3) of Lemma 19. $\square$

**Theorem 25 (Definitionally Equal Terms are Logically Related)**

1. If $\Gamma \vdash M = N : A$ then $\Gamma^- \vdash M = N \in [\![A^-]\!]$.

2. If $\Gamma \vdash A = B : K$ then $\Gamma^- \vdash A = B \in [\![K^-]\!]$.

**Proof:** Directly by Lemmas 23 and 24. $\square$

**Corollary 26 (Completeness of Algorithmic Equality)**

1. If $\Gamma \vdash M = N : A$ then $\Gamma^- \vdash M \Longleftrightarrow N : A^-$.

2. If $\Gamma \vdash A = B : K$ then $\Gamma^- \vdash A \Longleftrightarrow B : K^-$.

**Proof:** Directly by Theorem 25 and Theorem 19. $\square$

## 5 Soundness of Algorithmic Equality

In general, the algorithm for type-directed equality is not sound. However, when applied to valid objects of the same type, it is sound and relates only equal terms. This direction requires a number of lemmas established in Section 2.6, but is otherwise mostly straightforward.

**Lemma 27 (Subject Reduction)**
If $M \xrightarrow{\text{whr}} M'$ and $\Gamma \vdash M : A$ then $\Gamma \vdash M' : A$ and $\Gamma \vdash M = M' : A$.

**Proof:** By induction on the definition of weak head reduction, making use of the inversion and substitution lemmas.

**Case:**

$$\mathcal{W} = \frac{}{(\lambda x{:}A_1.\ M_2)\ M_1 \xrightarrow{\text{whr}} [M_1/x]M_2}$$

| | |
|---|---:|
| $\Gamma \vdash (\lambda x{:}A_1.\ M_2)\ M_1 : A$ | Assumption |
| $\Gamma \vdash \lambda x{:}A_1.\ M_2 : \Pi x{:}B_1.\ B_2$ | |
| $\Gamma \vdash M_1 : B_1$ | |
| $\Gamma \vdash [M_1/x]B_2 = A : \text{type}$ | By inversion (Lemma 9) |
| $\Gamma \vdash A_1 : \text{type}$ | |
| $\Gamma, x{:}A_1 \vdash M_2 : A_2$ | |
| $\Gamma \vdash \Pi x{:}A_1.\ A_2 = \Pi x{:}B_1.\ B_2 : \text{type}$ | By inversion (Lemma 9) |

$\Gamma \vdash A_1 = B_1 : \text{type}$
$\Gamma, x{:}A_1 \vdash A_2 = B_2 : \text{type}$   By injectivity of products (Lemma 12)
$\Gamma \vdash [M_1/x]M_2 : [M_1/x]A_2$   By substitution (Lemma 3)
$\Gamma \vdash [M_1/x]A_2 = [M_1/x]B_2 : \text{type}$   By substitution (Lemma 3)
$\Gamma \vdash [M_1/x]A_2 = A : \text{type}$   By transitivity
$\Gamma \vdash [M_1/x]M_2 : A$   By rule (type conversion)
$\Gamma \vdash A_1 : \text{type}$   Copied from above
$\Gamma, x{:}A_1 \vdash M_2 = M_2 : A_2$   By reflexivity
$\Gamma \vdash M_1 = M_1 : A_1$   By reflexivity
$\Gamma \vdash (\lambda x{:}A_1.M_2)\, M_1 = [M_1/x]M_2 : [M_1/x]A_2$   By rule (parallel conversion)
$\Gamma \vdash (\lambda x{:}A_1.M_2)\, M_1 = [M_1/x]M_2 : A$   By rule (type conversion)

**Case:**

$$\mathcal{W} = \dfrac{\begin{array}{c}\mathcal{W}_1 \\ M_1 \xrightarrow{\text{whr}} M_1'\end{array}}{M_1\, M_2 \xrightarrow{\text{whr}} M_1'\, M_2}$$

$\Gamma \vdash M_1\, M_2 : A$   Assumption
$\Gamma \vdash M_1 : \Pi x{:}A_2.\, A_1$
$\Gamma \vdash M_2 : A_2$
$\Gamma \vdash [M_2/x]A_1 = A : \text{type}$   By inversion (Lemma 9)
$\Gamma \vdash M_1' : \Pi x{:}A_2.\, A_1$   By i.h. on $\mathcal{W}_1$
$\Gamma \vdash M_1'\, M_2 : [M_2/x]A_1$   By rule (application)
$\Gamma \vdash M_1'\, M_2 : A$   By rule (type conversion)
$\Gamma \vdash M_1 = M_1' : \Pi x{:}A_2.\, A_1$   By inductive hypothesis
$\Gamma \vdash M_2 = M_2 : A_2$   By reflexivity
$\Gamma \vdash M_1\, M_2 = M_1'\, M_2 : [M_2/x]A_1$   By rule (simultaneous congruence)
$\Gamma \vdash M_1\, M_2 = M_1'\, M_2 : A$   By rule (type conversion)

$\square$

For the soundness of algorithmic equality we need subject reduction and validity (Lemma 7).

**Theorem 28 (Soundness of Algorithmic Equality)** *Assume $\Gamma$ is valid.*

1. *If $\Gamma \vdash M : A$ and $\Gamma \vdash N : A$ and $\Gamma^- \vdash M \Longleftrightarrow N : A^-$, then $\Gamma \vdash M = N : A$.*

2. *If $\Gamma \vdash M : A$ and $\Gamma \vdash N : B$ and $\Gamma^- \vdash M \longleftrightarrow N : \tau$, then $\Gamma \vdash M = N : A$, $\Gamma \vdash A = B : \text{type}$ and $A^- = B^- = \tau$.*

3. *If $\Gamma \vdash A : K$ and $\Gamma \vdash B : K$ and $\Gamma^- \vdash A \Longleftrightarrow B : K^-$, then $\Gamma \vdash A = B : K$.*

4. *If $\Gamma \vdash A : K$ and $\Gamma \vdash B : L$ and $\Gamma^- \vdash A \longleftrightarrow B : \kappa$, then $\Gamma \vdash A = B : K$, $\Gamma \vdash K = L : \text{kind}$ and $K^- = L^- = \kappa$.*

5. *If $\Gamma \vdash K : \text{kind}$ and $\Gamma \vdash L : \text{kind}$ and $\Gamma^- \vdash K \Longleftrightarrow L : \text{kind}^-$ then $\Gamma \vdash K = L : \text{kind}$.*

**Proof:** By induction on the structure of the given derivations for algorithmic equality, using validity and inversion on the typing derivations.

**Case:**

$$\mathcal{T} = \frac{x{:}\tau \text{ in } \Gamma^-}{\Gamma^- \vdash x \longleftrightarrow x : \tau}$$

| | |
|---|---:|
| $\Gamma \vdash x : A$ | Assumption |
| $\Gamma \vdash x : B$ | Assumption |
| $x{:}C$ in $\Gamma$, $\Gamma \vdash C = A :$ type, $\Gamma \vdash C = B :$ type | By inversion (Lemma 9) |
| $\Gamma \vdash A = B :$ type | By symmetry and transitivity |
| $\Gamma \vdash x = x : C$ | By rule |
| $\Gamma \vdash x = x : A$ | By type conversion |
| $A^- = B^- = C^- = \tau$ | By erasure preservation (Lemma 13) |

**Case:** $\mathcal{T}$ ends in an equality of constants. Like the previous case.

**Case:**

$$\mathcal{T} = \frac{\begin{array}{c}\mathcal{T}_1 \\ \Gamma^- \vdash M_1 \longleftrightarrow N_1 : \tau_2 \to \tau_1\end{array} \quad \begin{array}{c}\mathcal{T}_2 \\ \Gamma^- \vdash M_2 \Longleftrightarrow N_2 : \tau_2\end{array}}{\Gamma^- \vdash M_1\, M_2 \longleftrightarrow N_1\, N_2 : \tau_1}$$

| | |
|---|---:|
| $\Gamma \vdash M_1\, M_2 : A$ | Assumption |
| $\Gamma \vdash N_1\, N_2 : B$ | Assumption |
| $\Gamma \vdash M_1 : \Pi x{:}A_2.\ A_1$, | |
| $\Gamma \vdash M_2 : A_2$, and | |
| $\Gamma \vdash [M_2/x]A_1 = A :$ type | By inversion (Lemma 9) |
| $\Gamma \vdash \Pi x{:}A_2.\ A_1 :$ type | By validity (Lemma 7) |
| $\Gamma \vdash A_2 :$ type | |
| $\Gamma, x{:}A_2 \vdash A_1 :$ type | By inversion (Lemma 9) |
| $\Gamma \vdash N_1 : \Pi x{:}B_2.\ B_1$, | |
| $\Gamma \vdash N_2 : B_2$, and | |
| $\Gamma \vdash [N_2/x]B_1 = B :$ type | By inversion (Lemma 9) |
| $\Gamma \vdash \Pi x{:}B_2.\ B_1 :$ type | By validity (Lemma 7) |
| $\Gamma \vdash B_2 :$ type | |
| $\Gamma, x{:}B_2 \vdash B_1 :$ type | By inversion |
| $\Gamma \vdash M_1 = N_1 : \Pi x{:}A_2.\ A_1$, | |
| $\Gamma \vdash \Pi x{:}A_2.\ A_1 = \Pi x{:}B_2.\ B_1 :$ type, and | |
| $(\Pi x{:}A_2.\ A_1)^- = (\Pi x{:}B_2.\ B_1)^- = \tau_2 \to \tau_1$ | By i.h. on $\mathcal{T}_1$ |
| $\Gamma \vdash A_2 = B_2 :$ type and | |
| $\Gamma, x{:}A_2 \vdash A_1 = B_1 :$ type | By injectivity of products (Lemma 12) |
| $\Gamma \vdash N_2 : A_2$ | By symmetry and type conversion |
| $\Gamma \vdash M_2 = N_2 : A_2$ | By i.h. on $\mathcal{T}_2$ |
| $\Gamma \vdash M_1\, M_2 = N_1\, N_2 : [M_2/x]A_1$ | By rule |
| $\Gamma \vdash M_1\, M_2 = N_1\, N_2 : A$ | By type conversion |
| $\Gamma \vdash [M_2/x]A_1 = [N_2/x]B_1 :$ type | By family functionality |
| $A^- = A_1^- = B_1^- = B^- = \tau_1$ | By erasure preservation |

**Case:**

$$\mathcal{T} = \dfrac{\overset{\mathcal{W}}{M \xrightarrow{\text{whr}} M'} \quad \overset{\mathcal{T}'}{\Gamma^- \vdash M' \Longleftrightarrow N : P^-}}{\Gamma^- \vdash M \Longleftrightarrow N : P^-}$$

| | |
|---|---:|
| $\Gamma \vdash M : P$ | Assumption |
| $\Gamma \vdash N : P$ | Assumption |
| $\Gamma \vdash P : \text{type}$ | Validity (Lemma 7) |
| $\Gamma \vdash M' : P$ | By subject reduction (Lemma 27) |
| $\Gamma \vdash M' = N : P$ | By i.h. on $\mathcal{T}'$ |
| $\Gamma \vdash M = M' : P$ | By subject reduction (Lemma 27) |
| $\Gamma \vdash M = N : P$ | By transitivity |

**Case:** Reduction on the right-hand side follows similarly.

**Case:**

$$\mathcal{T} = \dfrac{\overset{\mathcal{S}}{\Gamma^- \vdash M \longleftrightarrow N : P^-}}{\Gamma^- \vdash M \Longleftrightarrow N : P^-}$$

| | |
|---|---:|
| $\Gamma \vdash M : P$ | Assumption |
| $\Gamma \vdash N : P$ | Assumption |
| $\Gamma \vdash M = N : P$ | By i.h. on $\mathcal{S}$ |

**Case:**

$$\mathcal{T} = \dfrac{\overset{\mathcal{T}_2}{\Gamma^-, x{:}\tau_1 \vdash M\,x \Longleftrightarrow N\,x : \tau_2}}{\Gamma^- \vdash M \Longleftrightarrow N : \tau_1 \to \tau_2}$$

| | |
|---|---:|
| $\Gamma \vdash M : \Pi x{:}A_1.\ A_2$ | Assumption |
| $\Gamma \vdash N : \Pi x{:}A_1.\ A_2$ | Assumption |
| $\Gamma \vdash \Pi x{:}A_1.\ A_2 : \text{type}$ | By assumption |
| $\Gamma \vdash A_1 : \text{type}$ | |
| $\Gamma, x{:}A_1 \vdash A_2 : \text{type}$ | Inversion |
| $A_1^- = \tau_1$ and $A_2^- = \tau_2$ | Assumption and definition of $()^-$ |
| $\Gamma, x{:}A_1 \vdash M\,x : A_2$ | By weakening and rule |
| $\Gamma, x{:}A_1 \vdash N\,x : A_2$ | By weakening and rule |
| $\Gamma, x{:}A_1 \vdash M\,x = N\,x : A_2$ | By i.h. on $\mathcal{T}_2$ |
| $\Gamma \vdash A_1 : \text{type}$ | By inversion (Lemma 9) |
| $\Gamma \vdash M = N : \Pi x{:}A_1.\ A_2$ | By extensionality rule |

□

**Corollary 29 (Logically Related Terms are Definitionally Equal)**
*Assume $\Gamma$ is valid.*

1. If $\Gamma \vdash M : A$, $\Gamma \vdash N : A$, and $\Gamma^- \vdash M = N \in [\![A^-]\!]$, then $\Gamma \vdash M = N : A$.

2. If $\Gamma \vdash A : K$, $\Gamma \vdash B : K$, and $\Gamma^- \vdash A = B \in [\![K^-]\!]$, then $\Gamma \vdash A = B : K$.

**Proof:** Direct from the assumptions and prior theorems. We show the proof for the first case.

| | |
|---|---:|
| $\Gamma^- \vdash M = N \in [\![A^-]\!]$ | Assumption |
| $\Gamma^- \vdash M \Longleftrightarrow N : A^-$ | By Theorem 19 |
| $\Gamma \vdash M = N : A$ | By Theorem 28 |

□

## 6 Decidability of Definitional Equality and Type-Checking

In this section we show that the judgment for algorithmic equality constitutes a decision procedure on valid terms of the same type. This result is then lifted to yield decidability of all judgments in the LF type theory.

The first step is to show that equality is decidable for terms that are algorithmically equal to themselves. Note that this property does not depend on the soundness or completeness of algorithmic equality—it is a purely syntactic result. The second step uses completeness of algorithmic equality and reflexivity to show that every well-typed term is algorithmically equal to itself. These two observations, together with soundness and completeness of algorithmic equality, yield the decidability of definitional equality for well-typed terms.

We say an object is *normalizing* iff it is related to some term by the type-directed equivalence algorithm. More precisely, $M$ is *normalizing* at simple type $\tau$ iff $\Delta \vdash M \Longleftrightarrow M' : \tau$ for some term $M'$. Note that by symmetry and transitivity of the algorithms, this implies that $\Delta \vdash M \Longleftrightarrow M : \tau$. A term $M$ is *structurally normalizing* iff it is related to some term by the structural equivalence algorithm. That is, $M$ is *structurally normalizing* iff $\Delta \vdash M \longleftrightarrow M' : \tau$ for some $M'$. A similar definition applies to families and kinds. Equality is decidable on normalizing terms.

**Lemma 30 (Decidability for Normalizing Terms)**

1. If $\Delta \vdash M \Longleftrightarrow M' : \tau$ and $\Delta \vdash N \Longleftrightarrow N' : \tau$ then it is decidable whether $\Delta \vdash M \Longleftrightarrow N : \tau$.

2. If $\Delta \vdash M \longleftrightarrow M' : \tau_1$ and $\Delta \vdash N \longleftrightarrow N' : \tau_2$ then it is decidable whether $\Delta \vdash M \longleftrightarrow N : \tau_3$ for some $\tau_3$.

3. If $\Delta \vdash A \Longleftrightarrow A' : \kappa$ and $\Delta \vdash B \Longleftrightarrow B' : \kappa$ then it is decidable whether $\Delta \vdash A \Longleftrightarrow B : \kappa$.

4. If $\Delta \vdash A \longleftrightarrow A' : \kappa_1$ and $\Delta \vdash B \longleftrightarrow B' : \kappa_2$ then it is decidable whether $\Delta \vdash A \longleftrightarrow B : \kappa_3$ for some $\kappa_3$.

5. If $\Delta \vdash K \Longleftrightarrow K' : \mathrm{kind}^-$ and $\Delta \vdash L \Longleftrightarrow L' : \mathrm{kind}^-$ then it is decidable whether $\Delta \vdash K \Longleftrightarrow L : \mathrm{kind}^-$.

**Proof:** We only sketch the proof of the first two properties—the others are similar. First note that $\Delta \vdash M \Longleftrightarrow N : \tau$ iff $\Delta \vdash M' \Longleftrightarrow N : \tau$ iff $\Delta \vdash M \Longleftrightarrow N' : \tau$ iff $\Delta \vdash M' \Longleftrightarrow N' : \tau$, so decidability of one implies decidability of the others with equal results. Given this observation, we prove parts (1) and (2) by simultaneous structural inductions on the given derivations. The critical lemma is the determinacy of algorithmic equality (Lemma 15). □

Now we can show decidability of equality via reflexivity and completeness of algorithmic equality.

**Theorem 31 (Decidability of Algorithmic Equality)** *Assume $\Gamma$ is valid.*

1. *If $\Gamma \vdash M : A$ and $\Gamma \vdash N : A$ then it is decidable whether $\Gamma^- \vdash M \Longleftrightarrow N : A^-$.*

2. *If $\Gamma \vdash A : K$ and $\Gamma \vdash B : K$ then it is decidable whether $\Gamma^- \vdash A \Longleftrightarrow B : K^-$.*

3. *If $\Gamma \vdash K : \mathrm{kind}$ and $\Gamma \vdash L : \mathrm{kind}$ then it is decidable whether $\Gamma^- \vdash K \Longleftrightarrow L : \mathrm{kind}^-$.*

**Proof:** We show only the proof of part (1) since the others are analogous.

By reflexivity of definitional equality (Lemma 2) and the completeness of algorithmic equality (Corollary 26), both $M$ and $N$ are normalizing. Hence by Lemma 30, algorithmic equivalence is decidable. □

**Corollary 32 (Decidability of Definitional Equality)** *Assume $\Gamma$ is valid.*

1. *If $\Gamma \vdash M : A$ and $\Gamma \vdash N : A$ then it is decidable whether $\Gamma \vdash M = N : A$.*

2. *If $\Gamma \vdash A : K$ and $\Gamma \vdash B : K$ then it is decidable whether $\Gamma \vdash A = B : K$.*

3. *If $\Gamma \vdash K : \mathrm{kind}$ and $\Gamma \vdash L : \mathrm{kind}$ then it is decidable whether $\Gamma \vdash K = L : K$.*

**Proof:** By soundness and completeness it suffices to check algorithmic equality which is decidable by Theorem 31. □

We now present an algorithmic version of type-checking that uses algorithmic equality as an auxiliary judgment. This is a purely bottom-up type-checker; more complicated strategies can also be justified with our results, but are beyond the scope of this paper.

**Objects**

$$\frac{x{:}A \text{ in } \Gamma}{\Gamma \vdash x \Rightarrow A} \qquad \frac{c{:}A \text{ in } \Sigma}{\Gamma \vdash c \Rightarrow A}$$

$$\frac{\Gamma \vdash M_1 \Rightarrow \Pi x{:}A_2'.\ A_1 \qquad \Gamma \vdash M_2 \Rightarrow A_2 \qquad \Gamma \vdash A_2' \Longleftrightarrow A_2 : \mathrm{type}}{\Gamma \vdash M_1\ M_2 \Rightarrow [M_2/x]A_1}$$

$$\frac{\Gamma \vdash A_1 \Rightarrow \mathrm{type} \qquad \Gamma, x{:}A_1 \vdash M_2 \Rightarrow A_2}{\Gamma \vdash \lambda x{:}A_1.\ M_2 \Rightarrow \Pi x{:}A_1.\ A_2}$$

**Families**

$$\frac{a \Rightarrow K \text{ in } \Sigma}{\Gamma \vdash a \Rightarrow K}$$

$$\frac{\Gamma \vdash A \Rightarrow \Pi x{:}B'.\ K \qquad \Gamma \vdash M \Rightarrow B \qquad \Gamma \vdash B' \Longleftrightarrow B : \mathrm{type}}{\Gamma \vdash A\ M \Rightarrow [M/x]K}$$

$$\frac{\Gamma \vdash A_1 \Rightarrow \mathrm{type} \qquad \Gamma, x{:}A_1 \vdash A_2 \Rightarrow \mathrm{type}}{\Gamma \vdash \Pi x{:}A_1.\ A_2 \Rightarrow \mathrm{type}}$$

**Kinds**

$$\frac{}{\Gamma \vdash \text{type} \Rightarrow \text{kind}} \qquad \frac{\Gamma \vdash A \Rightarrow \text{type} \quad \Gamma, x{:}A \vdash K \Rightarrow \text{kind}}{\Gamma \vdash \Pi x{:}A.\ K \Rightarrow \text{kind}}$$

Similar rules exist for checking validity of signatures and contexts.

**Lemma 33 (Correctness of Algorithmic Type-Checking)** *Assume $\Gamma$ is valid.*

1. *(Soundness) If $\Gamma \vdash M \Rightarrow A$ then $\Gamma \vdash M : A$.*

2. *(Completeness) If $\Gamma \vdash M : A$ then $\Gamma \vdash M \Rightarrow A'$ for some $A'$ such that $\Gamma \vdash A = A' : \text{type}$.*

**Proof:** Part 1 follows by induction on the structure of the algorithmic derivation, using validity (Theorem 7), soundness of algorithmic equality (Theorem 28) and the rule of type conversion.

Part 2 follows by induction on the structure of the typing derivation, using transitivity of equality, inversion on type equality, and completeness of algorithmic equality. □

**Theorem 34 (Decidability of Type-Checking)**

1. *It is decidable if $\Gamma$ is valid.*

2. *Given a valid $\Gamma$, $M$, and $A$, it is decidable whether $\Gamma \vdash M : A$.*

3. *Given a valid $\Gamma$, $A$, and $K$, it is decidable whether $\Gamma \vdash A : K$.*

4. *Given a valid $\Gamma$ and $K$, it is decidable whether $\Gamma \vdash K : \text{kind}$.*

**Proof:** Since the algorithmic typing rules are syntax-directed and algorithmic equality is decidable (Theorem 31), there either exists a unique $A'$ such that $\Gamma \vdash M \Rightarrow A'$ or there is no such $A'$. By correctness of algorithmic type-checking we then have $\Gamma \vdash M : A$ iff $\Gamma \vdash A' = A : \text{type}$, which is decidable by Theorem 32. □

The correctness of algorithmic type-checking also allows us to show strengthening, and a stronger form of the extensionality rule.

**Theorem 35 (Strengthening)** *For each judgment $J$ of the type theory, if $\Gamma, x{:}A, \Gamma' \vdash J$ and $x \notin \text{FV}(\Gamma') \cup \text{FV}(J)$, then $\Gamma, \Gamma' \vdash J$.*

**Proof:** Strengthening for the algorithmic version of type-checking follows by a simple structural induction, taking advantage of obvious strengthening for algorithmic equality. Strengthening for the original typing rules then follows by soundness and completeness of algorithmic typing. Strengthening for equality judgments follows from completeness (Corollary 26), soundness (Theorem 28), and strengthening for the typing judgment. □

**Corollary 36 (Strong Extensionality)** *The typing premises for $M$ and $N$ in the extensionality rule are redundant. That is, the following strong form of extensionality is admissible:*

$$\frac{\Gamma \vdash A_1 : \text{type} \quad \Gamma, x{:}A_1 \vdash M\ x = N\ x : A_2}{\Gamma \vdash M = N : \Pi x{:}A_1.\ A_2}$$

**Proof:** By inversion and strengthening.

| | |
|---|---:|
| $\Gamma, x{:}A_1 \vdash M\,x : A_2$ | By validity |
| $\Gamma, x{:}A_1 \vdash M : \Pi x{:}B_1.\,B_2,$ | |
| $\Gamma, x{:}A_1 \vdash x : B_1,$ and $\Gamma, x{:}A_1 \vdash B_2 = A_2 :$ type | By inversion (Lemma 9) |
| $\Gamma \vdash A_1 = B_1 :$ type | By inversion and strengthening |
| $\Gamma \vdash \Pi x{:}B_1.\,B_2 = \Pi x{:}A_1.\,A_2 :$ type | By rule |
| $\Gamma, x{:}A_1 \vdash M : \Pi x{:}A_1.\,A_2$ | By rule (type conversion) |
| $\Gamma \vdash M : \Pi x{:}A_1.\,A_2$ | By strengthening |
| $\Gamma \vdash N : \Pi x{:}A_1.\,A_2$ | Similarly |
| $\Gamma \vdash M = N : \Pi x{:}A_1.\,A_2$ | By extensionality |

□

## 7 Quasi-Canonical Forms

The representation techniques of LF mostly rely on compositional bijections between the expressions (including terms, formulas, deductions, *etc.*) of the object language and *canonical forms* in a meta-language, where canonical forms are $\eta$-long and $\beta$-normal forms. So if we are presented with an LF object $M$ of a given type $A$ and we want to know which object-language expression $M$ represents, we convert it to canonical form and apply the inverse of the representation function.

This leads to the question on how to compute the canonical form of a well-typed object $M$ of type $A$ in an appropriate context $\Gamma$. Generally, we would like to extract this information from a derivation that witnesses that $M$ is normalizing, that is, a derivation that shows that $M$ is algorithmically equal to itself. This idea cannot be applied directly in our situation, since a derivation $\Gamma^- \vdash M \Longleftrightarrow M : A^-$ yields no information on the type labels of the $\lambda$-abstractions in $M$. Fortunately, these turn out to be irrelevant: if we have an object $M$ of a given type $A$ which is in canonical form, possibly with the exception of some type labels, then the type labels are actually uniquely determined up to definitional equality.

We formalize this intuition by defining quasi-canonical forms (and the auxiliary notion of quasi-atomic forms) in which type-labels have been deleted. A quasi-canonical form can easily be extracted from a derivation that shows that a term is normalizing. Quasi-canonical forms are sufficient to prove adequacy theorems for the representation, since the global type of a quasi-canonical form is sufficient to extract an LF object unique up to definitional equality applied to type labels. The set of *quasi-canonical (QC)* and *quasi-atomic (QA)* terms are defined by the following grammar:

$$\begin{array}{lrcl}
\textit{Quasi-canonical objects} & \bar{\bar{M}} & ::= & \bar{M} \mid \lambda x.\,\bar{\bar{M}} \\
\textit{Quasi-atomic objects} & \bar{M} & ::= & x \mid c \mid \bar{M}\,\bar{\bar{M}}
\end{array}$$

It is a simple matter to instrument the algorithmic equality relations to extract a common quasi-canonical or quasi-atomic form for the terms being compared. Note that only one quasi-canonical form need be extracted, since two terms are algorithmically equivalent iff they have the same quasi-canonical form. The instrumented rules are as follows:

**Instrumented Type-Directed Object Equality**

$$\dfrac{M \xrightarrow{\text{whr}} M' \quad \Delta \vdash M' \Longleftrightarrow N : \alpha \Uparrow \bar{\bar{O}}}{\Delta \vdash M \Longleftrightarrow N : \alpha \Uparrow \bar{\bar{O}}} \qquad \dfrac{N \xrightarrow{\text{whr}} N' \quad \Delta \vdash M \Longleftrightarrow N' : \alpha \Uparrow \bar{\bar{O}}}{\Delta \vdash M \Longleftrightarrow N : \alpha \Uparrow \bar{\bar{O}}}$$

$$\dfrac{\Delta \vdash M \longleftrightarrow N : \alpha \downarrow \bar{O}}{\Delta \vdash M \Longleftrightarrow N : \alpha \Uparrow \bar{\bar{O}}} \qquad \dfrac{\Delta, x{:}\tau_1 \vdash M\,x \Longleftrightarrow N\,x : \tau_2 \Uparrow \bar{\bar{O}}}{\Delta \vdash M \Longleftrightarrow N : \tau_1 \to \tau_2 \Uparrow \lambda x.\,\bar{\bar{O}}}$$

**Instrumented Structural Object Equality**

$$\dfrac{x{:}\tau \text{ in } \Delta}{\Delta \vdash x \longleftrightarrow x : \tau \downarrow x} \qquad \dfrac{c{:}A \text{ in } \Sigma}{\Delta \vdash c \longleftrightarrow c : A^- \downarrow c}$$

$$\dfrac{\Delta \vdash M_1 \longleftrightarrow N_1 : \tau_2 \to \tau_1 \downarrow \bar{O}_1 \quad \Delta \vdash M_2 \Longleftrightarrow N_2 : \tau_2 \Uparrow \bar{\bar{O}}_2}{\Delta \vdash M_1\,M_2 \longleftrightarrow N_1\,N_2 : \tau_1 \downarrow \bar{O}_1\,\bar{\bar{O}}_2}$$

It follows from the foregoing development that every well-formed term has a unique quasi-canonical form. We now have the following theorem relating quasi-canonical forms to the usual development of the LF type theory. We write $|M|$ for the result of erasing the type labels from an object $M$.

**Theorem 37 (Quasi-Canonical and Quasi-Atomic Forms)**

1. If $\Gamma \vdash M_1 : A$ and $\Gamma \vdash M_2 : A$ and $\Gamma^- \vdash M_1 \Longleftrightarrow M_2 : A^- \Uparrow \bar{\bar{O}}$, then there is an $N$ such that $|N| = \bar{\bar{O}}$, $\Gamma \vdash N : A$, $\Gamma \vdash M_1 = N : A$ and $\Gamma \vdash M_2 = N : A$.

2. If $\Gamma \vdash M_1 : A_1$ and $\Gamma \vdash M_2 : A_2$ and $\Gamma^- \vdash M_1 \longleftrightarrow M_2 : \tau \Uparrow \bar{O}$ then $\Gamma \vdash A_1 = A_2 :$ type, $A_1^- = A_2^- = \tau$ and there is an $N$ such that $|N| = \bar{O}$, $\Gamma \vdash N : A_1$, $\Gamma \vdash M_1 = N : A_1$ and $\Gamma \vdash M_2 = N : A_1$.

**Proof:** By simultaneous induction on the instrumented equality derivations. It is critical that we have the types of the objects that are compared (and not just the approximate type) so that we can use this information to fill in the missing $\lambda$-labels. □

Note the $N$ in the theorem above is uniquely determined up to definitional equality of the type labels, since $\bar{\bar{O}}$ and $\bar{O}$ determine $N$ in all other respects. This result shows that all adequacy proofs for LF representation on canonical forms still hold. In fact, they can be carried out directly on quasi-canonical forms.

We can also directly state and prove prove adequacy theorems for encodings of logical systems in LF using quasi-canonical forms. It is interesting to observe that the type labels on $\lambda$'s are not necessary for this purpose; in an adequacy theorem, the type of the bound variable is determined from context. For example, the following relation sets up a compositional (natural) bijection between (a) terms and formulas of first-order logic over a given first-order signature and (b) quasi-canonical forms of types $\iota$ and $o$, respectively, in the signature of first-order logic. We only show an excerpt, illustrating the idea over the signature

$$\begin{aligned} c_f &: \iota \to \cdots \to \iota \\ c_= &: \iota \to \iota \to o \\ c_\wedge &: o \to o \to o \\ c_\forall &: (\iota \to o) \to o \end{aligned}$$

Let $\Gamma$ be a context of the form $x_1{:}\iota,\ldots,x_n{:}\iota$ for some $n \geq 0$. A correspondence relation between terms and formulas with (free) variables among the $x_1,\ldots,x_n$ and quasi-canonical objects of type $\iota$ and $o$, respectively, over that signature and context may be defined as follows:

$$\overline{\Gamma \vdash x \leftrightsquigarrow x : \iota}$$

$$\frac{\Gamma \vdash t_1 \leftrightsquigarrow \bar{\bar{M}}_1 : \iota \quad \ldots \quad \Gamma \vdash t_n \leftrightsquigarrow \bar{\bar{M}}_n : \iota}{\Gamma \vdash f(t_1,\ldots,t_n) \leftrightsquigarrow c_f\,\bar{\bar{M}}_1\,\ldots\,\bar{\bar{M}}_n : \iota} \quad \frac{\Gamma \vdash t_1 \leftrightsquigarrow \bar{\bar{M}}_1 : \iota \quad \Gamma \vdash t_2 \leftrightsquigarrow \bar{\bar{M}}_2 : \iota}{\Gamma \vdash t_1{=}t_2 \leftrightsquigarrow c_=\,\bar{\bar{M}}_1\,\bar{\bar{M}}_2 : o}$$

$$\frac{\Gamma \vdash \phi_1 \leftrightsquigarrow \bar{\bar{M}}_1 : o \quad \Gamma \vdash \phi_2 \leftrightsquigarrow \bar{\bar{M}}_2 : o}{\Gamma \vdash \phi_1 \wedge \phi_2 \leftrightsquigarrow c_\wedge\,\bar{\bar{M}}_1\,\bar{\bar{M}}_2 : o}$$

$$\frac{\Gamma, x{:}\iota \vdash \phi \leftrightsquigarrow \bar{\bar{M}} : o}{\Gamma \vdash \forall x.\,\phi \leftrightsquigarrow c_\forall\,(\lambda x.\,\bar{\bar{M}}) : o}$$

**Theorem 38 (Adequacy for Syntax of First-Order Logic)** *Let $\Gamma$ be a context of the form $x_1 : \iota, \ldots, x_n : \iota$ for some $n \geq 0$.*

1. *The relation $\Gamma \vdash t \leftrightsquigarrow M : \iota$ is a compositional bijection between terms $t$ of first-order logic over variables $x_1,\ldots,x_n$ and quasi-canonical forms $M$ of type $\iota$ relative to $\Gamma$.*

2. *The relation $\Gamma \vdash \phi \leftrightsquigarrow M : o$ is a compositional bijection between formulas $\phi$ with free variables among $x_1,\ldots,x_n$ and quasi-canonical forms $M$ of type $o$ relative to $\Gamma$.*

**Proof:** We establish by induction over the $t$ and $\phi$ that for every term $t$ and formula $\phi$ there exist a unique $M$ and $N$ and derivations of $\Gamma \vdash t \leftrightsquigarrow \bar{\bar{M}} : \iota$ and $\Gamma \vdash \phi \leftrightsquigarrow \bar{\bar{N}} : o$, respectively. Similarly, we show that for a quasi-canonical $\bar{\bar{M}}$ and $\bar{\bar{N}}$ at type $\iota$ and $o$, respectively, there exists unique related $t$ and $\phi$. This establishes a bijection. To see that it is compositional we use an induction over the structure of terms $t$ and formulas $\phi$. □

Adequacy at the level of derivations can be established by analogous means.

## 8 Conclusions

We have presented a new, type-directed algorithm for definitional equality in the LF type theory. This algorithm improves on previous accounts by avoiding consideration of reduction and its associated meta-theory and by providing a practical method for testing definitional equality in an implementation. The algorithm also yields a notion of canonical form, which we call quasi-canonical, that is suitable for proving the adequacy of encodings in a logical framework. The omission of type labels presents no difficulties for the methodology of LF, essentially because abstractions arise only in contexts where the domain type is known. The formulation of the algorithm and its proof of correctness relies on the "shapes" of types, from which dependencies on terms have been eliminated.

Surprisingly, it was the soundness proof for the algorithm, and not its completeness proof, that presented some technical difficulties. In particular, we have eliminated family-level $\lambda$-abstractions from our formulation of the type theory in order to prove injectivity of products syntactically.

The type-directed approach scales to richer languages such as those with unit types, products, and linear types [VC01], ordered types [PP99, Pol01], and proof irrelevant and intensional types [Pfe01] precisely because it makes use of type information during comparison. For example, one expects that any two variables of unit type are equal, even though they are structurally distinct head normal forms. A similar approach is used by Stone and Harper [SH00] to study a dependent type theory with singleton kinds and subkinding. There it is impossible to eliminate dependencies, resulting in a substantially more complex correctness proof, largely because of the loss of symmetry in the presence of dependencies. Nevertheless, the fundamental method is the same, and results in a practical approach to checking definitional equality for a rich type theory.

A major open question is if our technique be extended to handle the full Calculus of Constructions. We require injectivity of products rather early, which would seem to be difficult to attain. Furthermore, long normal forms, while still cleanly definable [DHW93], are not stable under substitutions which complicates the type-directed equality algorithm.

**Acknowledgments.** We are grateful to Chris Stone for improving the treatment of family functionality, and to Karl Crary and Jeff Polakow for several comments and corrections to an earlier draft of this paper. We would also like to thank the anonymous referees for a number of helpful suggestions and pointers to the literature.

# References

[Bar92]  Henk P. Barendregt. Lambda calculi with types. In S. Abramsky, D. Gabbay, and T.S.E. Maibaum, editors, *Handbook of Logic in Computer Science*, volume 2, chapter 2, pages 117–309. Oxford University Press, 1992.

[Cer96]  Iliano Cervesato. *A Linear Logical Framework*. PhD thesis, Dipartimento di Informatica, Università di Torino, February 1996.

[CG99]  Adriana Compagnoni and Healfdene Goguen. Antisymmetry of higher-order subtyping. In J. Flum and M. Rodríguez-Artalejo, editors, *Proceedings of the 8th Annual Conference on Computer Science Logic (CSL'99)*, pages 420–438, Madrid, Spain, September 1999. Springer-Verlag LNCS 1683.

[Coq91]  Thierry Coquand. An algorithm for testing conversion in type theory. In Gérard Huet and Gordon Plotkin, editors, *Logical Frameworks*, pages 255–279. Cambridge University Press, 1991.

[CP98]  Iliano Cervesato and Frank Pfenning. A linear logical framework. *Information and Computation*, 1998. To appear in a special issue with invited papers from LICS'96, E. Clarke, editor.

[DHW93]  Gilles Dowek, Gérard Huet, and Benjamin Werner. On the definition of the eta-long normal form in type systems of the cube. In Herman Geuvers, editor, *Informal Proceedings of the Workshop on Types for Proofs and Programs*, Nijmegen, The Netherlands, May 1993.

[FM90]  Amy Felty and Dale Miller. Encoding a dependent-type λ-calculus in a logic programming language. In M.E. Stickel, editor, *10th International Conference on Automated Deduction*, pages 221–235, Kaiserslautern, Germany, July 1990. Springer-Verlag LNCS 449.

[GB99]   Herman Geuvers and Erik Barendsen. Some logical and syntactical observations concerning the first order dependent type system λP. *Mathematical Structures in Computer Sciences*, 9(4):335–360, 1999.

[Geu92]  Herman Geuvers. The Church-Rosser property for $\beta\eta$-reduction in typed $\lambda$-calculi. In A. Scedrov, editor, *Seventh Annual IEEE Symposium on Logic in Computer Science*, pages 453–460, Santa Cruz, California, June 1992.

[Gha97]  Neil Ghani. Eta-expansions in dependent type theory — the calculus of constructions. In P. de Groote and J.R. Hindley, editors, *Proceedings of the Third International Conference on Typed Lambda Calculus and Applications (TLCA'97)*, pages 164–180, Nancy, France, April 1997. Springer-Verlag LNCS 1210.

[Gog94]  Healfdene Goguen. *A Typed Operational Semantics for Type Theory*. PhD thesis, Edinburgh University, August 1994.

[Gog99]  Healfdene Goguen. Soundness of the logical framework for its typed operational semantics. In Jean-Yves Girard, editor, *Proceedings of the 4th International Conference on Typed Lambda Calculi and Applications (TLCA'99)*, pages 177–197, L'Aquila, Italy, April 1999. Springer-Verlag LNCS 1581.

[HHP93]  Robert Harper, Furio Honsell, and Gordon Plotkin. A framework for defining logics. *Journal of the Association for Computing Machinery*, 40(1):143–184, January 1993.

[HP99]   Robert Harper and Frank Pfenning. On equivalence and canonical forms in the LF type theory. Technical Report CMU-CS-99-159, Department of Computer Science, Carnegie Mellon University, 1999.

[JG95]   C. B. Jay and N. Ghani. The virtues of $\eta$-expansion. *Journal of Functional Programming*, 5(2):135–154, 1995.

[NPS90]  B. Nordström, K. Petersson, and J.M. Smith. *Programming in Martin-Löf's Type Theory: An Introduction*. Oxford University Press, 1990.

[Pfe92]  Frank Pfenning. Computation and deduction. Unpublished lecture notes, 277 pp. Revised May 1994, April 1996, May 1992.

[Pfe93]  Frank Pfenning. Refinement types for logical frameworks. In Herman Geuvers, editor, *Informal Proceedings of the Workshop on Types for Proofs and Programs*, pages 285–299, Nijmegen, The Netherlands, May 1993.

[Pfe01]  Frank Pfenning. Intensionality, extensionality, and proof irrelevance in modal type theory. In J. Halpern, editor, *Proceedings of the 16th Annual Symposium on Logic in Computer Science (LICS'01)*, pages 221–230, Boston, Massachusetts, June 2001. IEEE Computer Society Press. Extended version available as Technical Report CMU-CS-01-116, Department of Computer Science, Carnegie Mellon University, September 2001.

[Pol01]  Jeff Polakow. *Ordered Linear Logic and Applications*. PhD thesis, Department of Computer Science, Carnegie Mellon University, August 2001.

[PP99]   Jeff Polakow and Frank Pfenning. Natural deduction for intuitionistic non-commutative linear logic. In J.-Y. Girard, editor, *Proceedings of the 4th International Conference on Typed Lambda Calculi and Applications (TLCA'99)*, pages 295–309, L'Aquila, Italy, April 1999. Springer-Verlag LNCS 1581.

[Sal90]   Anne Salvesen. The Church-Rosser theorem for LF with $\beta\eta$-reduction. Unpublished notes to a talk given at the First Workshop on Logical Frameworks in Antibes, France, May 1990.

[SH00]   Christopher A. Stone and Robert Harper. Decidable type equivalence with singleton kinds. In Thomas Reps, editor, *Conference Record of the 27th Symposium on Principles of Programming Languages (POPL'00)*, pages 214–227, Boston, Massachusetts, January 2000. ACM Press.

[Str91]   Thomas Streicher. *Semantics of Type Theory*. Birkhäuser, 1991.

[VC01]   Joseph C. Vanderwaart and Karl Crary. A simplified account of the metatheory of linear LF. Technical Report CMU-CS-01-154, Carnegie Mellon University, 2001.

[Vir99]   Roberto Virga. *Higher-Order Rewriting with Dependent Types*. PhD thesis, Department of Mathematical Sciences, Carnegie Mellon University, September 1999. Available as Technical Report CMU-CS-99-167.